\documentclass[12pt]{article}%
\usepackage[switch,columnwise]{lineno}
\usepackage{amsmath}%
\usepackage[dvips]{color}
\usepackage{txfonts}%
\usepackage{amsfonts}%
\usepackage{amssymb}%
\usepackage{graphicx}
\usepackage{color}
\usepackage{setspace}
\usepackage{esint}
\usepackage[rightcaption]{sidecap}
\usepackage{wrapfig}
\usepackage{lipsum}
\usepackage{enumitem}
\setlist{nosep}
\usepackage{caption}
\usepackage{longtable}
\usepackage{multirow}
\usepackage[T2A,T1]{fontenc}
\usepackage[utf8]{inputenc}
\usepackage[russian,english]{babel}

\usepackage{chngcntr}

\numberwithin{equation}{section}

\newcommand{\rr}{\color{red}}
\newcommand{\be}{\begin{equation}}
\newcommand{\ee}{\end{equation}}

\usepackage{anysize}
\marginsize{2.5cm}{2.5cm}{1cm}{2cm}
\usepackage{mathptmx}
-1in\relax 
\begin{document}
\begin{center}
\LARGE
\textbf{Energy Balance Within Thermonuclear Reactors}
\normalsize
\vskip1cm
\begin{tabular}{ll}
\Large
\Large Mikhail V. Shubov         \\
\normalsize
University of MA Lowell    \\
One University Ave,        \\
Lowell, MA 01854           \\
E-mail: mikhail\_shubov@uml.edu  \\
\end{tabular}
\end{center}
%\hskip1.7cm*\emph{Corresponding Author.}  E-mail: viktor\_shubov@uml.edu

%\footnotetext[2010]{\textbf{Mathematics Subject Classification:} 34N99; 65Z05; 76K05; 80A20}
%\blfootnote{\textbf{Keywords:} projectile; maximum range; impact energy; aerodynamic drag loss; aerodynamic heating.}

\begin{quote}
\begin{quote}
\begin{center}
  \textbf{Abstract}
\end{center}
Thermonuclear reactors hold a great promise for the future of Humankind. Within Tokamak and Stellarator reactors, plasma is confined by twisted magnetic fields. Reactors which produce fusion energy have existed since Princeton Large Torus Tokamak in 1978, nevertheless in all reactors built up to now, energy loss from plasma vastly exceeded fusion energy production. In order for a thermonuclear power plant to run, generated fusion energy must significantly exceed energy loss by the plasma. There are four processes by which plasma looses energy -- neutron radiation, Bremsstrahlung radiation, synchrotron radiation, and heat conduction to the walls. For a deuterium -- tritium reactor, 80\% of energy produced by fusion is lost to neutron radiation, about 4\% to 6\% of fusion energy is lost to Bremsstrahlung and synchrotron radiation. For a deuterium -- $^3$He reactor, 5\% of energy produced by fusion is lost to neutron radiation, about 50\% to 75\% of fusion energy is lost to Bremsstrahlung and synchrotron radiation. Increasing reactor operating temperature decreases loss to Bremsstrahlung radiation and increases loss to synchrotron radiation. Power loss to conduction is either independent or weakly dependent on fusion power production. Up to now, no single theoretical or experimental model can accurately predict conduction power loss. For small Tokamaks and Stellarators, conduction power loss vastly exceeds power generated by fusion. For large and powerful thermonuclear reactors which are yet to be built, conduction power loss should be much lower than power produced by fusion.
\end{quote}
\end{quote}

\large
\begin{center}
  \textbf{List of Notations}
\end{center}
\normalsize
\begin{center}
  \textbf{Latin Notations}
\end{center}
\begin{description}
  \item[$a$] -- Tokamak or Stellarator minor radius
  \item[$A=R/a$] -- Tokamak or Stellarator parameter where $a$ is the minor radius and $R$ is the major radius
  \item[$a_i$] -- molar abundances of nuclei in plasma
  \item[$B$] -- the magnetic field
  \item[$B_{_{\text{critical}}}(T)$] -- temperature-dependent critical magnetic field for a superconductor
  \item[$B_{_T}$] -- toroidal field at the plasma major radius in a vacuum shot
  \item[$B_{_P}$] -- poloidal field at plasma surface
  \item[$C_{_{Ld}}$] -- Lawson criterion expressed in terms of particle density
  \item[$C_{_{LP}}$] -- \textbf{Lawson pressure criterion}, which is Lawson Criterion expressed in terms of pressure
  \item[$C_{_{LPS}}$] -- Lawson pressure criterion for Stellarator reactor
  \item[$C_{_{LPT}}$] -- Lawson pressure criterion for Tokamak reactor
  \item[$C_{_{LPTh}}$] -- theoretically expected Lawson pressure criterion
  \item[$f_{_B}$] -- the fraction of power produced by nuclear fusion in plasma, which is carried off as Bremsstrahlung radiation
  %\item[$F_{_{\text{Fuel}}}$] -- factor determined by reactor fuel
  \item[$f_{_{\text{heat}}}$] -- the fraction of fusion power which is used to heat the plasma exclusive of the power lost to Bremsstrahlung and synchrotron radiation
  \item[$f_{_{\text{n}}}$] -- neutronicity or the fraction of fusion power carried away by neutrons
  %\item[$F_{_{\text{Reactor}}}$] -- factor determined by the intrinsic characteristics of the reactor -- size, shape, and maximum magnetic field
  %\item[$F_{_{\text{RO}}}$] -- factor determined by the reactor operating conditions
  \item[$f_{_{S\kappa}}\big(\kappa,A\big)$] -- proportionality function which determines dependence of Stellarator Lawson pressure criterion on plasma elongation
  \item[$f_{_{T\kappa}}\big(\kappa,A\big)$] -- proportionality function which determines dependence of Tokamak Lawson pressure criterion on plasma elongation
  \item[$f_{_{\text{Synchrotron}}}$] -- fraction of fusion power radiated away as synchrotron radiation exclusive of the power reflected by the wall and reabsorbed by plasma
  \item[$f_{_{\gamma}}$] -- fraction of fusion power radiated away via synchrotron and Bremsstrahlung radiation
  \item[$\overline{g}_B$] -- frequency average of the speed averaged Gaunt factor
  \item[$H$] -- the ratio of the actual and predicted confinement times
  \item[$I$] -- toroidal current flowing through the plasma in Tokamak or Stellarator
  \item[$M$] -- mean plasma ion mass in $amu$
  \item[$n$] -- ion number density in plasma
  \item[$n_e$] -- electron density in plasma
  \item[$n_{_G}$] -- Greenwald density limit
  \item[$P$] -- heating power of the plasma in Tokamak or Stellarator
  \item[$P_{_{\text{Conductive}}}$] -- conduction power loss by Tokamak or Stellarator plasma
  \item[$\mathcal{P}$] -- fusion power per unit volume
  \item[$\mathcal{P}_{_B}$] -- Bremsstrahlung radiation power per unit of volume
  \item[$P_{_{_{\text{fusion}}}}$] -- total fusion power of a Tokamak or Stellarator
  \item[$\mathcal{P}_{_{\text{fh}}}$] -- fusion power per unit volume used to heat plasma
  \item[$P_{_g}$] -- plasma gas pressure
  \item[$P_{_G}$] -- plasma pressure limit based on Greenwald density limit
  \item[$\mathcal{P}_{_{\text{heat}}}$] -- external heating power per unit volume needed to heat plasma is
  \item[$\mathcal{P}_{_{\text{loss}}}$] -- energy loss of plasma per unit volume by conduction
  \item[$P_{_m}$] -- magnetic field pressure
  \item[$P_{_{\text{Tokamak}}}$] -- conduction power loss by Tokamak plasma
  \item[$P_{_{\text{Stellarator}}}$] -- conduction power loss by Stellarator plasma
  \item[$\mathcal{P}_{_{\text{Synchrotron}}}$] --  synchrotron radiation power per unit volume
  \item[$q=1/\iota$] -- safety factor, where $\iota$ is the rotational transform of the field lines within the Tokamak or Stellarator torus
  \item[$q_{_{95}}$] -- plasma edge safety factor
  \item[$r$] -- nuclear reaction rate in $m^{-3}s^{-1}$
  \item[$R$] -- Tokamak or Stellarator major radius
  \item[$\mathcal{R}_{_C}=S_{_F}^{3} B_{_T}^{4} R^{3}$] -- Reactor Criterion, which is the main criterion determining if the reactor produces more energy than it consumes or vice versa
  \item[$R_{_G}$] -- \textbf{Greenwald ratio} is the ratio of plasma pressure to pressure limit based on Greenwald density
  \item[$S_{_F}$] -- shape factor
  \item[$\hat{S}_{_F}$] -- scaled shape factor $\hat{S}_{_F}=S_{_F}\ A$
  \item[$T$] -- temperature
  \item[$T_{_{\text{critical}}}$] -- critical temperature for a superconductor
  \item[$V_{_g}$] -- plasma volume
  \item[$W$] -- thermal energy of plasma
  \item[$\mathcal{W}$] is kinetic energy density of plasma
  \item[$w_{_r}$] -- wall reflectivity of synchrotron radiation
  \item[$x_{_1}$] -- proportion of deuterium in plasma by the number of nuclei
  \item[$x_{_2}$] -- is the proportion of either tritium or $^3$He in plasma by the number of nuclei
  \item[$x_{_b}$] -- proportion of the fuel burned
  \item[$x_{_i}$] -- proportion of impurities within plasma
  \item[$x_{_r}$] -- constant defined by
      \[
      x_{_r}=4\ x_{_1}\ x_{_2}.
      \]
  \item[$Z$] -- the average nuclear charge
 \end{description}

\begin{center}
  \textbf{Greek Notations}
\end{center}

\begin{description}
  \item[$\alpha_{_{\text{Coefficient}}}$] -- power of a coefficient given in a model for energy confinement time presented in (\ref{1.04.12})
  \item[$\beta$] -- the quotient of plasma pressure and magnetic field pressure
  \item[$\beta_{_N}$] -- normalized beta
  \item[$\delta$] -- plasma triangularity
  \item[$\eta$] -- electric resistivity of plasma
  \item[$\iota$] -- \textbf{magnetic helicity} or rotational transform of the field lines within the Tokamak or Stellarator torus
  \item[$\kappa$] -- plasma elongation, defined as the quotient of the plasma volume and the volume of torus with minor radius $a$ and major radius $R$
      \[
      \kappa=\frac{V}{2 \pi^2 R a^2}
      \]
  \item[$\sigma_{_{P}}(T)$] -- temperature-dependent power production rate parameter defined in terms of pressure and measured in $bar^{-2} W m^{-3}$
  \item[$\sigma_{_{v}}(T)$] -- temperature-dependent reaction rate parameter defined in terms of number density and  measured in $m^3/s$
  \item[$\sigma_{_{B}}(T)$] -- burning rate constant given by
      \[
        \frac{1}{n} \frac{dn}{dt}=-\sigma_{_{B}}(T)\ P_{_g}
      \]
  \item[$\tau_{_E}$] --  energy confinement time
  \item[$\tau_{_{ET}}$] --  energy confinement time for Tokamak or Stellarator reactors
  \item[$\tau_{_{ES}}$] --  energy confinement time for Stellarator reactors
\end{description}

\begin{center}
  \textbf{Constants}
\end{center}
\begin{description}
  \item[$k$] -- the Boltzmann constant:
     \[
     k=1.381 \cdot 10^{-23} \frac{J}{^oK}
     \]
  \item[$R_{_{g}}$] -- the gas constant:
    \[
    R_{_{g}}=8.314 \frac{J}{mol \ ^oK}
            =1.6015 \cdot 10^{-21} \frac{bar \  m^3 }{keV}
    \]
  \item[$\varepsilon_0$] -- electric constant:
    \[
    \varepsilon_0=\frac{1}{\mu_{_0}\ c^2}=8.85 \cdot 10^{-12} \frac{F}{m}
    \]
  \item[$\mu_{_0}$] -- magnetic constant or vacuum permeability:
    \[
    \mu_{_0}=4 \pi \cdot 10^{-7}\ \frac{N}{A^2}
    \]
\end{description}

\section{Introduction}
Thermonuclear fusion holds a great promise as the power source for Human Civilization. Energy production is one of the primary driving forces for industrialization and economic progress of Humankind \cite{ShubovPM}.  In this work we discuss possibilities, advances and technical challenges in the field of sustainable nuclear fusion power.  Thermonuclear energy has a potential of becoming almost unlimited power source on Earth.  Further, in the future, thermonuclear energy can provide an almost unlimited power source during the colonization of the Solar System.

The resources of thermonuclear fuel on Earth are truly vast.
Earth's oceans contain 52 trillion tons of deuterium \cite[p.10]{Fus1}.
Even though tritium does not exist in nature, it is generated from $^6$Li isotope.
Lithium contains 6.6\% isotope $^6$Li \cite[p.1-15]{crc}.
World Lithium reserves are 14 million tons in ores \cite[p.99]{minerals2019} and 180 billion tons in seawater \cite[p.14-14]{crc}.  Lithium contained in Earth's oceans can be used to generate 5.9 billion tons of tritium.

World energy reserves of fossil fuel are equal to 0.93 trillion ton oil equivalent, while possible resources are 12.6 trillion ton oil equivalent \cite{WorldEnergy}.
Thermonuclear reactors using deuterium -- tritium fusion can generate an energy equivalent of $7.5\cdot 10^4$ trillion tons of oil from tritium obtained from oceanic lithium.
Reactors using pure deuterium fusion can generate an energy equivalent of $3.3\cdot 10^8$ trillion tons of oil from 52 trillion tons of oceanic deuterium.

Within the Solar System, thermonuclear fuel is even more abundant.  Each ton of Jupiter's atmosphere contains about 100 $g$ deuterium and 11 $g$ $^3$He \cite{Jupiter_He3}.  Thus, Jupiter contains $18 \cdot 10^{18}\ tons$ $^3$He and $170 \cdot 10^{18}\ tons$ deuterium.  As we discuss in this work, building a reactor using deuterium -- $^3$He fusion may be difficult, yet these difficulties should be overcome by the time when the Outer Solar System is being colonized.

The primary fusion reaction of interest in this century is deuterium -- tritium fusion
    \be
    \label{1.01.01}
    \text{$^2$H+$^3$H$\ \to ^4$He+n+17.6}\ MeV.
    \ee
Even though tritium does not exist in nature, it can be produced from the isotope $^6$Li via the reaction
    \be
    \label{1.01.02}
    \text{$^6$Li+n$\ \to ^4$He+$^3$H}.
    \ee
Energetic neutrons produced by fusion knock out neutrons of other atoms before being slowed down and consumed by $^6$Li.  On average, one nuclear fusion event, which consumes one tritium atom, produces 1.2 tritium atoms \cite[p.67]{Triton1}.

A fusion reaction which does not produce neutrons is called aneutronic.  The only aneutronic fusion reaction which can be run in a Tokamak or Stellarator is deuterium -- $^3$He fusion
    \be
    \label{1.01.03}
    \text{$^2$H+$^3$He$\ \to ^4$He+p+18.5}\ MeV
    \ee
Other aneutronic fusion reactions exist, but they are impractical to run in Tokamaks and Stellarators.  Deuterium -- $^3$He reactors would require much more advanced technology than deuterium -- tritium reactors.  Moreover, resources of $^3$He on Earth are miniscule, thus $^3$He would have to be mined in space \cite{EarthHe3}.  Deuterium -- $^3$He reactors should appear long after deuterium -- tritium reactors have reached maturity.

Fusion reaction can occur only in a plasma at very high temperatures of 100 -- 200 million $^o$K \cite{Tok2}.  The temperature of plasma is measured in kiloelectronvolts ($keV$):
      \be
      \label{1.01.04}
      1\ keV=1.16 \cdot 10^7\ ^oK.
      \ee
Plasma at this temperature can be confined either by inertia or by a magnetic field.  In this work we consider only magnetic confinement.

In Toakamak and Stellarator reactors, plasma confined within a magnetic field has a shape of torus.  Original reactors built in the 1950s were faced with plasma leakage problem.  Due to unevenness of toroid field, charged particles would drift out of confinement.
In Tokamak reactors, this problem is solved by induction of toroidal electric current through the plasma.  This current generates a poloidal magnetic field.  Overall magnetic field lines take a shape of a twisted torus.  The term Tokamak originated in Russian language in USSR -- Toroidalnya Kamera Magnetnaya Katushka \cite[p.2]{Tok2}.
In Stellarator reactors, the problem of plasma drifting is solved by introducing a poloidal magnetic field by "careful profiling of the magnetic field topology through complex
non-planar toroidal field coils" \cite[p.32]{Tokamak}.  Stellarator magnetic fields are very complex with up to 50 features which have to be accounted for in design \cite[p. 31]{Fus2}.
A Tokamak reactor is illustrated in Figure \ref{1.0F01} below.
\begin{center}
\includegraphics[width=12cm,height=12cm]{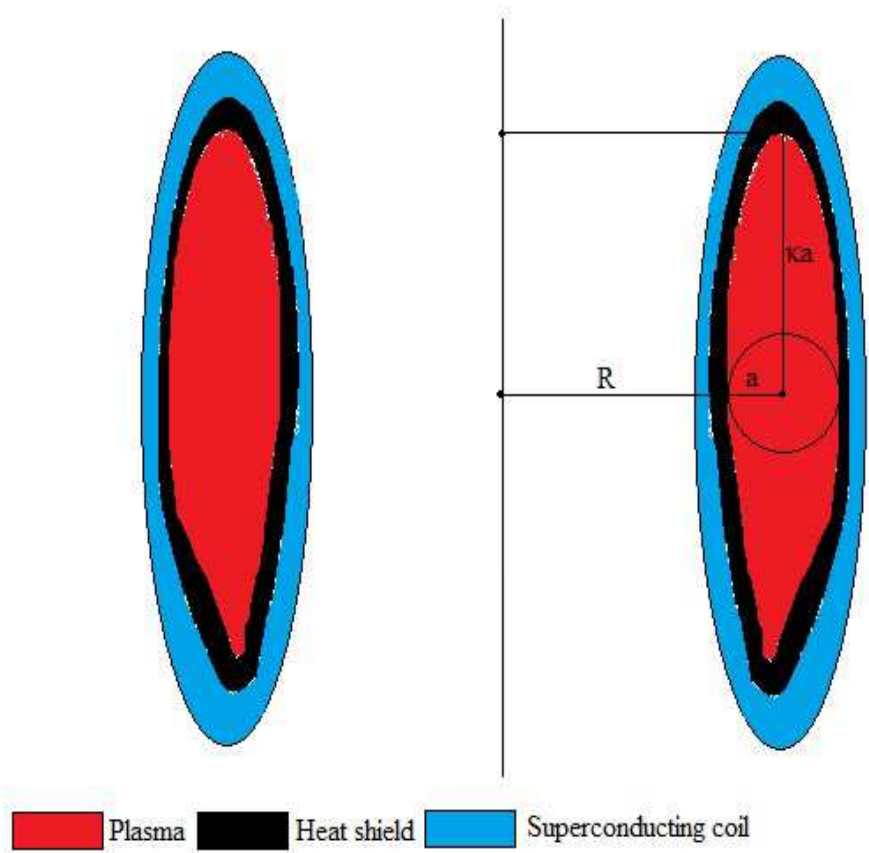}
\captionof{figure}{Tokamak cross-section \label{1.0F01}}
\end{center}

The first Tokamak was built in 1952 in Kurchatov Institute, Moscow, Russia \cite[p. 20]{Fus3}.  Confinement of deuterium-tritium plasma at temperature of 5 $keV$ has been achieved at Princeton Large Torus Tokamak in 1978 \cite[p. 78]{Hist02}.  This was the first Tokamak to produce fusion energy.

Even though Tokamaks and Stellarators which produce fusion energy have been built, the energy produced in these reactors is much lower than the energy consumed by external plasma heating.  Tokamak or stellarator plasma rapidly loses energy by four processes --  neutron radiation, Bremsstrahlung radiation, synchrotron radiation, and heat conduction to the walls.  Bremsstrahlung radiation is caused by collision of energetic electrons with nuclei.  It consists of x-rays.  Synchrotron radiation is caused by radiation of electrons moving in the magnetic field.  It consists of visible and infrared photons.  Conduction heat loss is caused by interaction of the plasma with the walls.

Plasma energy loss by neutron, Bremsstrahlung radiation and synchrotron radiation are proportional to energy generated by fusion.  As we show in Subsection {\rr 4.1}, in a typical deuterium-tritium reactor, 80\% of fusion energy escapes plasma with neutrons, 6.7\% with Bremsstrahlung radiation and 1.3\% with synchrotron radiation.  As we show in Subsection {\rr 4.2}, conductive energy loss is either independent or weakly dependent on the fusion power.  Big ITER reactor designed in great detail (not built) has conductive power loss of 182 $MW$
\cite[p. 7]{BigITER}.  Lowest conductive power loss for any designed Tokamak is about 40 $MW$ \cite[p. 10]{STCensus01}.

In order for the confined plasma to generate more energy than it loses, the plasma must have sufficient pressure and \textbf{energy confinement time} \cite[p.2]{Tok2}.  Energy confinement time $\tau_{_E}$ is the ratio of plasma internal heat energy and the rate of energy loss by the plasma \cite[p.18]{Tokamak}.  Notice, that energy confinement time is not related to the time during which the plasma itself is confined \cite[p.4]{Lawson1}.  Any working fusion power plant must have stable plasma.  The product of plasma pressure and confinement time is called Lawson pressure criterion and denoted $C_{_{LP}}$.
As we show in Subsection {\rr 4.1}, in order for deuterium -- tritium fusion reactor to operate, it must have\\ $C_{_{LP}} \ge 16\ bar \cdot s$.
A deuterium -- $^3$He fusion reactor must have
$C_{_{LP}} \ge 430\ bar \cdot s$ \cite[p.81]{AF01}.

As we demonstrate in Subsection {\rr 4.2}, the Lawson pressure criterion of a reactor is proportional to its reactor criterion defined as
    \be
    \label{1.01.05}
    \mathcal{R}_{_C}=S_{_F}^{3} B_{_T}^{4} R^{3}.,
    \ee
where $R$ is the major radius of the plasma torus, $B_{_T}$ is the toroidal magnetic field on the plasma axis, and $S_{_F}$ is the shape factor defined in Subsection {\rr 4.2}.  In order for a thermonuclear reactor to sustain nuclear fusion, it must have sufficiently high reactor criterion.  Reactor criterion threshold depends on reactor shape, toroidal current, and especially on the fuel used.

For any reactor with shape similar to International Thermonuclear Experimental Reactor (ITER), the thresholds are the following:
    \be
    \label{1.01.06}
    \begin{split}
    \mathcal{R}_{_C}\Big(\ ^2\text{H}-^3\text{H} \Big)=3.2 \cdot 10^7\ Tesla^4\ m^3, \qquad
    \mathcal{R}_{_C}\Big(\ ^2\text{H}-^3\text{He} \Big)=1.4 \cdot 10^{10}\ Tesla^4\ m^3.
    \end{split}
    \ee
Two proposed deuterium -- $^3$He spherical Tokamaks discussed in Subsection {\rr 4.6} have reactor criteria of about $5 \cdot 10^{10}\ Tesla^4\ m^3$.  ITER obtains sufficient reactor criterion by having a major plasma radius $R=6.2\ m$, toroidal magnetic field $B_{_T}=5.3\ Tesla$, and shape factor $S_{_F}=4.2$ \cite[p.2]{EConf03}.  Increasing toroidal magnetic field presents significant technical challenges, which are likely to be solved within coming decades.

Compact fusion reactors may be developed in the future.  They will achieve sufficient reactor criterion by increasing the toroidal magnetic field \cite{HField1}.  Achieving high $B_{_T}=5.3\ Tesla$ is difficult -- as we show in Table \ref{1.0T12}, the maximum field experienced by the superconducting coils is 2 to 3 times higher than $B_{_T}$.  The development of high temperature superconductor tape should significantly increase reactor magnetic fields \cite{HTSTape03,SCMS01,HTSTape04}.

In Section 2, we briefly introduce the physics of Tokamak and Stellarator plasmas.
In Subsection 2.1, we derive expressions for plasma pressure, magnetic field pressure, and $\beta$ -- the ratio of the two pressures.
In Subsection 2.2, we derive the expression for electrical resistivity of the plasma, and show that it is a good conductor.
In Subsection 2.3, we derive the nuclear reaction rate within plasma.

In Section 3, we describe and calculate radiative energy losses from plasma.
In Subsection 3.1, we describe the four types of plasma energy loss -- neutron radiation, Bremsstrahlung radiation, synchrotron radiation, and conduction to walls.
In Subsection 3.2, we derive the energy loss due to Bremsstrahlung radiation.
In Subsection 3.3, we derive the energy loss due to synchrotron radiation.
Energy losses from both Bremsstrahlung and synchrotron radiation are significant only for deuterium -- $^3$He reactors.

In Section 4, we discuss transport energy loss.
In Subsection 4.1, we introduce the concept of \textbf{fusion power gain} $Q$, which is the ratio of thermal energy produced by fusion to heating power which must be supplied to the plasma in order to sustain fusion.  We also introduce \textbf{Lawson pressure criterion} -- which is the product of plasma pressure and energy confinement time.
In Subsections 4.2 and 4.3, we analyze scaling laws for energy confinement time and conductive energy loss by plasma.
In Subsections 4.3 to 4.5, we define the reactor criterion and calculate it for several reactors.

In Section 5, we discuss the ways in which reactor criterion and Lawson pressure criterion can be improved.
In Subsection 5.1, we introduce the Greenwald density limit and show that it does place a flexible upper bound on size and power of reactors.
In Subsection 5.2, we discuss the simple approach of increasing a Tokamak of Stellarator major radius.
In Subsection 5.3, we discuss advantages and disadvantages of using spherical Tokamaks.
In Subsection 5.4, we discuss technology which can increase the magnetic field which confines plasma.
In Subsection 5.5, we discuss advantages and disadvantages of running a reactor in an unsafe regime.

In Section 6, we discuss prospects for reactor development.
In Subsection 6.1, we discuss a possible timeline for nuclear fusion reactors.
In Subsection 6.2, we describe Spheromak2100 -- a concept of a future deuterium -- $^3$He reactor.

\section{Tokamak and Stellerator plasma}
In this section we introduce the physics of plasma present in Tokamaks and Stellarators.  This plasma is very hot and completely ionized.  The physics of this plasma is different from that of "low temperature" or partially ionized plasmas.

Plasma physics consists of many dynamic and electric phenomena which have been subject to extensive studies \cite{Tokamak}.  In this section, we only consider the aspects of plasma physics most relevant to operation of Tokamaks and Stellarators.

\subsection{Plasma pressure and $\beta$}

Plasma pressure is an important parameter of Tokamak or Stellerator plasma.  As we demonstrate below, the power of a thermonuclear reactor is proportional to the square of plasma pressure.  The magnetic field required to contain plasma is proportional to the square root of plasma pressure.

Plasma Pressure is calculated in the same way as the pressure of any monatomic gas.  The particle density of plasma is the total number of ions and electrons per unit volume.  It is $(1+\overline{Z})\ n$ where $\overline{Z}$ is the average ion charge and $n$ is the number density of electrons.  Thus, by the laws of perfect gas, the plasma pressure is \cite[p.45]{Fus1}:
    \be
    \label{1.02.01}
    P_{_g}=(1+\overline{Z})\ n\ R_{_{g}}\ T,
    \ee
where the gas constant $R_{_{g}}$ is
    \be
    \label{1.02.02}
    R_{_{g}}=8.314 \frac{J}{mol \cdot ^oK}=1.6015 \cdot 10^{-21} \frac{bar \cdot m^3 }{keV}.
    \ee
The notation $R$, generally used for gas constant, is reserved for the Tokamak's or Stellarator's major radius.  Substituting (\ref{1.02.01}) into (\ref{1.02.02}), we obtain the pressure of plasma:
    \be
    \label{1.02.03}
    P_{_{g}}=1.6015 \cdot 10^{-21} \frac{bar \cdot m^3 }{keV}\ (1+\overline{Z})\ n\ T.
    \ee
Subscript $g$ denotes gas.

Magnetic field pressure is
    \be
    \label{1.02.04}
    P_{_m}=\frac{B^2}{2 \mu_{_0}}=3.98\ bar \left(\frac{B}{1\ Tesla}\right)^2.
    \ee
The quotient of plasma pressure and magnetic field pressure is
\cite[p.34]{Fus1}:
    \be
    \label{1.02.05}
    \beta^*=\frac{P_{_g}}{P_{_m}}.
    \ee
Given that both plasma pressure and magnetic field are variable over space, we define $\beta$ as the average given by \cite[p.17]{Tk02}
    \be
    \label{1.02.06}
    \beta=\frac{2 \mu_{_0} \sqrt{\int P_{_g}^2(V)\ dV}}{B_{_T}^2},
    \ee
where $V$ is the plasma volume and $B_{_T}$ is the toroidal magnetic field within a Tokamak or Stellerator \cite[p.17]{Tk02}.

For most Tokamaks, $\beta$ is usually close to 0.01, or 1\% \cite[p.30]{Fus3}.  International Thermonuclear Experimental Reactor (ITER) will have $\beta=2.4$\% \cite[p.26]{Tk03}  Spherical Tokamaks have values of $\beta$ up to 40\% \cite[p.333]{Tokamak}.  Spherical Tokamaks should be more compact than conventional ones \cite{ST04}.

\subsection{Electrical resistivity of plasma}

In Tokamaks, toroidal currents of 15 $MA$ and more would have to flow through the plasma  \cite[p.226]{FusBk1}.  In order for the current to be stable, the plasma should be a strong electric conductor.

The specific electric resistivity of plasma is
    \be
    \label{1.02.07}
    \eta=\frac{\sqrt{m_{_e}}\ \overline{Z}\ e^2\ \ln \Lambda}{91.5 \epsilon_{_{0}}^2}
    \big(k T_{_e} \big)^{-3/2},
    \ee
where $T_{_e}$ is the electron temperature, and
 $\ln \Lambda \approx 20$ is the Coulomb logarithm.
For hydrogen isotope plasma \cite[p.19]{FusBk2},
    \be
    \label{1.02.08}
    \eta=3.3 \cdot 10^{-8}\ \left( \frac{T}{1\ keV} \right)^{-3/2}\
    \big( \Omega \cdot m \big).
    \ee
The specific resistivity of copper at 20$^o$C is
$1.8 \cdot 10^{-8}\ \big( \Omega \cdot m \big)$.
In a typical fusion reactor which uses deuterium -- tritium reaction, the temperature is 10 $keV$ and the plasma resistivity is
$1.0 \cdot 10^{-9}\ \Omega \cdot m$, which is 18 times lower than copper resistivity.
In a typical fusion reactor which uses deuterium -- $^3$He reaction, the temperature is 45 $keV$, the average nuclear charge is $\overline{Z}=1.5$ and the plasma resistivity is
$1.6 \cdot 10^{-10}\ \Omega \cdot m$, which is 110 time lower than copper resistivity.

\subsection{Nuclear reaction rate within plasma}

A two component plasma consists of two types of nuclei.  The reaction rate in a two-component plasma is
    \be
    \label{1.02.09}
    r=n_{_{1}} n_{_{2}}\sigma_{_{v}}(T),
    \ee
where $n_{_{1}}$ and $n_{_{2}}$ are number densities of reactant nuclei and $\sigma_{_{v}}(T)$ is the temperature-dependent reaction rate parameter measured in $m^3/s$ .  In a reactor plasma,
    \be
    \label{1.02.10}
    \begin{split}
    n_{_{1}}&=x_{_1}\ n\\
    n_{_{2}}&=x_{_2}\ n,
    \end{split}
    \ee
where $x_{_1}$ is the proportion of deuterium in plasma,
$x_{_2}$ is the proportion of either tritium or $^3$He in plasma,
and $1-x_{_1}-x_{_2}$ is the proportion of reaction products and impurities in plasma.

Denote
    \be
    \label{1.02.11}
    x_{_r}=4\ x_{_1}\ x_{_2}.
    \ee
The maximum value of $x_{_r}$ is 1.  This maximum is reached if the two reactants are present in equal proportion, and no fuel has been consumed yet.
The reaction rate is
    \be
    \label{1.02.12}
    r=\frac{x_{r}}{4}\ n^2\ \sigma_{_{v}}(T),
    \ee
where $n$ is the number density of all nuclei \cite[p.38]{Fus1}.

Substituting (\ref{1.02.01}) into (\ref{1.02.12}), we obtain the following reaction rate
    \be
    \label{1.02.13}
    r=
    \frac{x_{_r}\ \sigma_{_{v}}(T)}{4\ (1+\overline{Z})\ R_{_{g}}\ T}\ P_{_g}\ n=
    \frac{x_{_r}\ \sigma_{_{v}}(T)}{4\ (1+\overline{Z})^2\ R_{_{g}}^2\ T^2}\ P_{_g}^2.
    \ee
From (\ref{1.02.13}) above, we deduce the fuel consumption rate
    \be
    \label{1.02.14}
    \frac{1}{n} \frac{dn}{dt}=-
    \frac{x_{_r}\ \sigma_{_{v}}(T)}{4\ (1+\overline{Z})\ R_{_{g}}\ T}\ P_{_g}=-
    \frac{x_{_r}}{2}\ \sigma_{_{B}}(T)\ P_{_g},
    \ee
where
    \be
    \label{1.02.15}
    \sigma_{_{B}}(T)=
    \frac{\sigma_{_{v}}(T)}{2\ (1+\overline{Z})\ R_{_{g}}\ T}
    \ee
is the temperature-dependent burning rate constant.

The fusion power per unit volume of two component plasma is given by
    \be
    \label{1.02.16}
    \mathcal{P}=
    \frac{x_{_r}}{4}\ n^2\ \sigma_{_{v}}(T)E_{_r}=
    \frac{x_{_r}\ \sigma_{_{v}}(T) E_{_r} }{4\ (1+\overline{Z})^2\ R_{_{g}}^2\ T^2}\ P_{_g}^2=
    \frac{x_{_r}}{4}\ \sigma_{_{P}}(T)\ P_{_g}^2,
    \ee
where $E_{_r}$ is the energy released by a single nuclear reaction.
The \textbf{specific power constant} is
    \be
    \label{1.02.17}
    \sigma_{_{P}}(T)=
    \frac{\sigma_{_{v}}(T) E_{_r} }{(1+\overline{Z})^2\ R_{_{g}}^2\ T^2}.
    \ee
The reaction
    \be
    \label{1.02.18}
    ^2H+^3H \to ^4He+n
    \ee
has $E_{_r}=2.82 \cdot 10^{-12}\ J$ \cite[p.40]{Fus1}.

Substituting (\ref{1.02.04}) and (\ref{1.02.06}) into (\ref{1.02.16}) we obtain
    \be
    \label{1.02.19}
    \mathcal{P}=15.8\ bar^2 \ x_{_r}\
    \sigma_{_{P}}(T)\ \left(\frac{B_{_T}}{1\ Tesla}\right)^4 \beta^2
    \ee
for deuterium -- tritium plasma.  The values of $\sigma_{_{v}}(T)$ for deuterium -- tritium plasma are tabulated below \cite[p.46]{Fus1}.
The value of $\sigma_{_{B}}(T)$ is obtained by (\ref{1.02.15}).
The value of $\sigma_{_{P}}(T)$ is obtained by (\ref{1.02.17}).
The units of $\sigma_{_{P}}(T)$ is $bar^{-2}  W  m^{-3}$, which can be rewritten as
    \be
    \label{1.02.20}
    \begin{split}
    &1\ bar^{-2}  W  m^{-3}=1\ bar^{-2}\ s^{-1}\ \big( J\ m^{-3}\big)=1\ bar^{-2}\ s^{-1}\ Pa\\
    =&1\ bar^{-2}\ s^{-1}\ Pa\ 10^{-5}\ bar=10^{-5}\ bar^{-1}\ s^{-1}.
    \end{split}
    \ee
\begin{center}
  \begin{tabular}{|l|l|l|l|l|l|l|l|l|l|l|l|}
     \hline
     Quantity & Unit & & & & & & & & & \\
     \hline
     T &$keV$                              & 5    & 8    & 10   & 15   & 20   & 25  & 30  & 35  & 40 \\
     \hline
     $\sigma_{_{v}}(T)$ & $10^{-22}\ m^3 s^{-1}$ &
     0.13 & 0.59 & 1.09 & 2.65 & 4.24 & 5.6 & 6.7 & 7.5 & 8.0\\
     \hline
     $\sigma_{_{B}}(T)$
     & $10^{-3}\ bar^{-1}  s^{-1}$ &
     %0.20 & 0.58 &0.85 &1.38 &1.65 &1.75 &1.74 &1.67 &1.56\\
     0.41 & 1.15 & 1.7 &2.76 &3.31 &3.5  &3.49 &3.35 &3.12\\
     \hline
     $\sigma_{_{P}}(T)$ &
     $bar^{-1}  s^{-1}$ &
     %0.36 & 0.63 & 0.75 & 0.81 & 0.73 & 0.62 & 0.51 & 0.42 & 0.34 \\
     1.43  & 2.53 &3.00  &3.24  &2.91  &2.46  &2.05  &1.68  & 1.37 \\
     \hline
   \end{tabular}
   \captionof{table}{Reaction rate parameter for deuterium-tritium fusion} \label{1.0T01}
\end{center}

Similar parameters for deuterium -- $^3$He reaction are tabulated below.  Cross-section data \cite[p.10]{RRates0} is used to calculate the last two rows.
\begin{center}
  \begin{tabular}{|l|l|l|l|l|l|l|l|l|l|l|l|}
     \hline
     Quantity & Unit & & & & & & & & & & \\
     \hline
     T &$keV$
     & 30 & 40   &  50   & 60   & 70   &  80  & 90  & 100  & 120 & 140\\
     \hline
     $\sigma_{_{v}}(T)$ & $10^{-23}\ m^3 s^{-1}$ &
     %0.033 & 0.072 &
     0.13 & 0.31 & 0.54  & 0.79 & 1.04 & 1.27 & 1.48& 1.67 & 1.97& 2.19 \\
     \hline
     $\sigma_{_{B}}(T)$
     & $10^{-5}\ bar^{-1}  s^{-1}$ &
     %0.10 & 0.18 & 0.27 &  0.48 &  0.67 &  0.82 &  0.93 &  0.99 &  1.03 &  1.03\\
      0.54 & 0.97 & 1.35 &  1.64 &  1.86 &  1.98 &  2.05 &  2.09 &  2.05 &  1.95\\
     \hline
     $\sigma_{_{P}}(T)$ &
     $10^{-2}\ bar^{-1}  s^{-1}$ &
    %0.36 & 0.51 & 0.64  & 0.85 &  0.95 &  0.97 &  0.93 &  0.87 &  0.80 &  0.6 \\
     2.54 & 3.41 & 3.80  & 3.86 &  3.73 &  3.49 &  3.21 &  2.94 &  2.41 &  1.97 \\
     \hline
   \end{tabular}
   \captionof{table}{Reaction rate parameter for deuterium-$^3$He fusion} \label{1.0T02}
\end{center}

Peak specific power constants for several fusion reactions are tabulated below.  The data for Column 1 is given in \cite{RRates}.  Column 2 is measured in units of
$10^{-27}\ m^3\ s^{-1}\ keV^2$.
\begin{center}
  \begin{tabular}{|l|r|c|c|c|l|l|l|l|l|l|l|l|l|}
  \hline
  Reaction              & Maximal             & $\overline{Z}$ & $E_{_r}$      & Maximal $\sigma_{_{P}}$       & Temperature \\
                        & $\sigma_{_{v}}/T^2$ &                & $10^{-12}\ J$ & $bar^{-1}  s^{-1}$            & $keV$       \\

  \hline
  $^2$H+$^3$H  $ \to ^4$He+n  & 1,240 & 1.0 & 2.82 & 3.4                & 13.6        \\
  $^2$H+$^3$He $ \to ^4$He+p  & 22.4  & 1.5 & 2.93 & $4.1 \cdot 10^{-2}$ & 58          \\
  p+$^6$Li $ \to ^4$He+$^3$He & 1.5   & 2.0 & 0.64 & $9.2 \cdot 10^{-4}$ & 66          \\
  p+$^{11}$B $ \to 3 ^4$He    & 3.0   & 3.0 & 1.39 & $4.0 \cdot 10^{-3}$ & 123         \\
  \hline
  \end{tabular}
  \captionof{table}{Specific power constant} \label{1.0T03}
\end{center}
As we see from the table above, deuterium -- $^3$He reaction is the only aneutronic fusion reaction worth considering in the foreseeable future.

\subsection{Calculation of $x_r$}
In this subsection, we calculate $x_r$ for deuterium -- tritium and deuterium -- $^3$He plasmas.  Suppose we start out with plasma with equal concentration of deuterium and the second component.  The proportion of fuel which has burned is $x_{_b}$.  Obviously, $x_{_b}$ changes over time of a Tokamak discharge.  The proportion of impurity nuclei is $x_{_i}$.  Impurity proportion should be under 2\% and it can be assumed to be static.

First, we calculate $x_{_b}$ for deuterium -- tritium reactor plasma.
Prior to fusion, the number density of deuterium and tritium nuclei is
    \be
    \label{1.02.21}
    n_{_1}=n_{_2}=\frac{n \big(1-x_{_i}\big)}{2}.
    \ee
The number density of deuterium and tritium nuclei after $x_{_b}$ of fuel has burned is
    \be
    \label{1.02.22}
    n_{_1}=n_{_2}=\frac{n \big(1-x_{_i}\big)\big(1-x_{_b} \big)}{2}.
    \ee
Deuterium -- tritium fusion transforms two fuel nuclei into one $^4$He nucleus.  Thus, the number density of products is
    \be
    \label{1.02.23}
    n_{_{\text{Products}}}=\frac{n \big(1-x_{_i}\big) x_{_b}}{2}.
    \ee
The number density of impurities is
    \be
    \label{1.02.24}
    n_{_i}=n x_{_i}.
    \ee
Adding (\ref{1.02.22}), (\ref{1.02.23}), and (\ref{1.02.24}) we obtain the overall number density of particles after $x_{_b}$ of fuel has been consumed:
    \be
    \label{1.02.25}
    n=n_{_1}+n_{_2}+n_{_{\text{Products}}}+n_{_i}=
    n\left[1-\frac{x_{_b}\big(1-x_{_i} \big)}{2} \right].
    \ee
Dividing (\ref{1.02.22}) by (\ref{1.02.25}) we obtain
    \be
    \label{1.02.26}
    x_{_1}=x_{_2}=\frac{ \big(1-x_{_i}\big)\big(1-x_{_b} \big)}
    {2-x_{_b}\big(1-x_{_i} \big)}.
    \ee
Given that $x_{_i} \ll 1$, we simplify (\ref{1.02.26}):
    \be
    \label{1.02.27}
    \begin{split}
    x_{_1}=x_{_2}&=\frac{ \big(1-x_{_b} \big)\big(1-x_{_i}\big)}
    {2-x_{_b}\big(1-x_{_i} \big)}=
    \frac{ \big(1-x_{_b} \big)\big(1-x_{_i}\big)}
    {\big(2-x_{_b}\big)\left[1-\frac{x_{_i}\ x_{_b}}{2-x_{_b}}\right]}
    \approx
    \frac{1-x_{_b}}{2-x_{_b}}\big(1-x_{_i} \big)
    \left[1-\frac{x_{_i}\ x_{_b}}{2-x_{_b}}\right]\\
    &
    \approx
    \frac{1-x_{_b}}{2-x_{_b}}
    \left[1-x_{_i}\left(1+ \frac{x_{_b}}{2-x_{_b}}\right) \right]=
    \frac{1-x_{_b}}{2-x_{_b}}
    \left[1-\frac{2 x_{_i}}{2-x_{_b}} \right].
    \end{split}
    \ee
Substituting (\ref{1.02.27}) into (\ref{1.02.11}), we obtain $x_{_r}$ for deuterium -- tritium reactor plasma:
    \be
    \label{1.02.28}
    x_{_r}=4 \left[\frac{1-x_{_b}}{2-x_{_b}}\right]^2
    \left[1-\frac{2 x_{_i}}{2-x_{_b}} \right]^2
    \approx
    \left[\frac{2-2\ x_{_b}}{2-x_{_b}}\right]^2
    \left[1-\frac{4 x_{_i}}{2-x_{_b}} \right].
    \ee

Second, we calculate $x_{_b}$ for deuterium -- $^3$He reactor plasma.  Prior to fusion, the number density of deuterium and $^3$He nuclei is also given by (\ref{1.02.21}).  The number density of deuterium and $^3$He nuclei after $x_{_b}$ of fuel has burned is still given by (\ref{1.02.22}).  Deuterium -- $^3$He fusion transforms two fuel nuclei into two product nuclei.  The total number density of nuclei remains unchanged in the process of deuterium -- $^3$He fusion.  The number density of particles after $x_{_b}$ of fuel has been consumed is $n$ -- the same as the original number density.  Dividing (\ref{1.02.22}) by $n$ we obtain
    \be
    \label{1.02.29}
    x_{_1}=x_{_2}=\frac{ \big(1-x_{_i}\big)\big(1-x_{_b} \big)}{2}.
    \ee
Substituting (\ref{1.02.29}) into (\ref{1.02.11}), we obtain $x_{_r}$ for deuterium -- $^3$He reactor plasma:
    \be
    \label{1.02.30}
    x_{_r}=\big(1-x_{_b}\big)^2\big(1-x_{_i}\big)^2
    \approx
    \big(1-x_{_b}\big)^2\ \big(1-2x_{_i}\big).
    \ee

In calculations presented above, we have ignored other nuclear reactions taking place within the plasma.  Deuterium -- deuterium fusion will have relatively minor effect on plasma composition.  Other fusion reactions would have negligible effect.

\section{Radiative energy loss}
\subsection{Reactor energy balance}
Reactors in which deuterium-tritium fusion took place have existed since 1978.  This has been achieved by confining deuterium-tritium plasma at fusion temperature within a Tokamak or Stellarator torus \cite[p. 78]{Hist02}.  The principal problem for past and present Tokamaks and Stellarators is the fact that energy produced by fusion has been much lower than energy lost by plasma.  Thus, in order to sustain fusion, thermal energy has to be supplied to the plasma by an external source.  In thermonuclear reactors, the plasma has to be heated with beams of energetic neutral particles or electromagnetic waves of radio or microwave frequencies \cite[p. 64-70]{FusBk1}.

In order for a fusion power station to operate, heat produced by fusion must considerably exceed heat supplied to the plasma.  Plasma in which heat generation by fusion exceeds energy loss does not need an external heat source to sustain fusion.  This plasma is called \textbf{ignited}.

There are four processes by which heat escapes plasma.  First is neutron radiation.  Deuterium -- tritium fusion given by $^2$H+$^3$H --> $^4$He+n releases 80\% of energy with the neutron.  Deuterium -- deuterium fusion takes place in deuterium -- $^3$He reactors.  About half of this fusion is given by $^2$H+$^2$H --> $^3$He+n.  This reaction releases 75\% of energy with the neutron.  Overall, 80\% of energy generated by deuterium-tritium fusion and 5\% of energy generated by deuterium -- $^3$He fusion is released as neutron radiation \cite[p.24]{Tokamak}.

The second form of energy loss is Bremsstrahlung radiation resulting from collisions of charged particles.  As we show in Subsection {\rr 3.2} below,  Bremsstrahlung energy loss decreases with increasing reactor operating temperature.

The third form of energy loss is Synchrotron radiation.  This radiation is caused by charged particles moving in a magnetic field.  This form of energy loss is significant only for deuterium -- $^3$He reactors.  Most Synchrotron radiation is absorbed by the plasma, with only 1.5\% to 4.5\% reaching the reactor wall \cite[p.70]{BeegITER}.  Theoretical calculations on the total power loss by Synchrotron radiation are vague, with different theories giving results different by up to a factor of 2 \cite[p.70]{BeegITER}.  As we show in Subsection {\rr 3.3}, energy loss by Synchrotron radiation increases with increasing reactor operating temperature.

Energy loss for Bremsstrahlung radiation increases with lowering of reactor operating temperature, while energy loss for Synchrotron radiation increases with increasing reactor operating temperature.  As we show in Table \ref{1.0T13} of Subsection {\rr 6.2}, the ideal operating temperature for deuterium -- $^3$He reactors is 60 $keV$ to 70 $keV$.

The fourth form of energy loss is conduction or transport.  Plasma confined in the magnetic field still fills up the volume of reactor torus -- thus the plasma touches reactor walls.  The behavior of plasma is not completely understood, thus many experimental and theoretical models exist for plasma energy loss via conduction \cite[p.138]{FusBk3}.  Conduction energy loss is very important for both deuterium -- tritium and deuterium -- $^3$He reactors.

Energy losses in the form of neutron, Bremsstrahlung, and synchrotron radiation are proportional to energy generated by nuclear fusion.  Neutronic energy loss is easiest to calculate.  Bremsstrahlung energy loss is somewhat easy to calculate.  Calculation of synchrotron energy loss is complicated, and different theories produce results differing by  up to a factor of 2 \cite[p. 70]{BeegITER}.  In this section, we calculate plasma energy loss via three types of radiation -- neutron, Bremsstrahlung, and synchrotron.

Energy losses in the form of conduction is weakly related or unrelated to energy generated by nuclear fusion.  This mechanism of energy loss is not completely understood, thus there are different physical and empirical models.  This form of energy loss is discussed in Section {\rr 4}.

\subsection{Bremsstrahlung radiation}
Fully ionized plasma produces electromagnetic radiation via Bremsstrahlung -- radiation resulting from collisions of charged particles.  The total power per unit volume is \cite[p.162]{bslr}:
    \be
    \label{1.03.01}
    \begin{split}
    \mathcal{P}_{_B}&=1.4 \cdot 10^{-40}  \frac{W}{m^3}
    \cdot
    \left( \frac{T}{1 ^oK}\right)^{1/2}
    \left( \frac{n_e}{m^{-3}}\right)^2
    \left( \frac{\sum a_i Z_i^2}{\sum a_i Z_i} \right)\overline{g}_B,\\
    \end{split}
    \ee
where $\overline{g}_B$ ``is a frequency average of the speed averaged Gaunt factor,
which is in the range 1.1 to 1.5. Choosing a value of 1.2 will give an
accuracy to within about 20\%" \cite[p.162]{bslr}.  Substituting (\ref{1.02.03}) and into (\ref{1.03.01}) we obtain
    \be
    \label{1.03.02}
    \begin{split}
    \mathcal{P}_{_B}&=1.9 \cdot 10^5  \frac{W}{m^3}
    \cdot
    \left( \frac{T}{1\ keV}\right)^{-3/2}
    \left( \frac{P_{_g}}{1\ bar}\right)^2
    \left( \frac{\overline{Z}\ \overline{Z^2}}{\big(1+\overline{Z}\big)^2} \right)\overline{g}_B,\\
    \end{split}
    \ee
where overline denotes average value.

Dividing (\ref{1.03.01}) by (\ref{1.02.16}), we obtain the fraction of fusion power radiated away via Bremsstrahlung radiation.  This fraction is called \textbf{Bremsstrahlung power fraction}:
    \be
    \label{1.03.03}
    \begin{split}
    f_{_B}&=
    \frac{\mathcal{P}_{_B}}{\mathcal{P}}=
    \frac{7.6 \cdot 10^5  \frac{W}{m^3\cdot bar^2}}{x_{_r}\ \sigma_{_{P}}(T)}
    \left( \frac{T}{1\ keV}\right)^{-3/2}
    \left( \frac{\overline{Z}\ \overline{Z^2}}{\big(1+\overline{Z}\big)^2} \right)\overline{g}_B\\
    &=\frac{7.6\ bar^{-1} s^{-1}}{x_{_r}\ \sigma_{_{P}}(T)}
    \left( \frac{T}{1\ keV}\right)^{-3/2}
    \left( \frac{\overline{Z}\ \overline{Z^2}}{\big(1+\overline{Z}\big)^2} \right)\overline{g}_B.
    \end{split}
    \ee
As we see from (\ref{1.03.03}) above, the Bremsstrahlung power fraction does not depend on the plasma pressure.

We take $\overline{g}_B=1.2$.  For deuterium-tritium fusion with 20\% of fuel burned, $\overline{Z}=1.11$, and\\ $\overline{Z^2}=1.33$.  Substituting these values into (\ref{1.03.03}), we obtain
    \be
    \label{1.03.04}
    f_{_B}=
    \frac{3.0 \ bar^{-1} s^{-1}}{x_{_r}\ \sigma_{_{P}}(T)}
    \left( \frac{T}{1\ keV}\right)^{-3/2}.
    \ee

Using the data presented in Table \ref{1.0T01}, we calculate the Bremsstrahlung power fraction of deuterium -- tritium plasma as a function of temperature.  This fraction is tabulated in Table \ref{1.0T04} below:
\begin{center}
  \begin{tabular}{|l|l|l|l|l|l|l|l|l|l|l|l|}
     \hline
     Quantity & Unit & & & & & & & & & \\
     \hline
     T &$keV$
     & 5    & 8    & 10   & 15   & 20   & 25  & 30  & 35  & 40 \\
     \hline
     $\sigma_{_{P}}(T)$ &
     $bar^{-1}  s^{-1}$ &
     1.43  & 2.53 &3.00  &3.24  &2.91  &2.46  &2.05  &1.68  & 1.37 \\
     \hline
     $x_{_r}\ f_{_B}$ &
     &0.19 &  0.052 &  0.032 &  0.016  & 0.012 &  0.010 &  0.009 &  0.009 &  0.009\\
     \hline
   \end{tabular}
   \captionof{table}{Bremsstrahlung power fraction of deuterium -- tritium plasma} \label{1.0T04}
\end{center}

Taking $\overline{g}_B=1.2$, $\overline{Z}=1.5$, and $\overline{Z^2}=2.5$ for deuterium -- $^3$He fusion reactor we obtain the following
    \be
    \label{1.03.05}
    f_{_B}\Big(\ ^2\text{H}-^3\text{He} \Big)=
    \frac{5.5 \ bar^{-1} s^{-1}}{x_{_r}\ \sigma_{_{P}}(T)}
    \left( \frac{T}{1\ keV}\right)^{-3/2}.
    \ee
Using the data presented in Table \ref{1.0T02}, we calculate the Bremsstrahlung power fraction of deuterium -- $^3$He plasma as a function of temperature.  This fraction is tabulated in Table \ref{1.0T05} below:
\begin{center}
  \begin{tabular}{|l|l|l|l|l|l|l|l|l|l|l|l|}
     \hline
     Quantity & Unit & & & & & & & & & & \\
     \hline
     T &$keV$
     & 30 & 40   &  50   & 60   & 70   &  80  & 90  & 100  & 120 & 140\\
     \hline
     $\sigma_{_{P}}(T)$ &
     $10^{-2}\ bar^{-1}  s^{-1}$ &
     2.54 & 3.41 & 3.80  & 3.86 &  3.73 &  3.49 &  3.21 &  2.94 &  2.41 &  1.97 \\
     \hline
     $x_{_r}\ f_{_B}$ & &
     1.32 & 0.64 & 0.41  & 0.31 &  0.25 &  0.22 &  0.2  & 0.19  & 0.17  & 0.17 \\
     \hline
   \end{tabular}
   \captionof{table}{Bremsstrahlung Power Fraction of Deuterium -- $^3$He Plasma} \label{1.0T05}
\end{center}
In all fusion reactions other than deuterium -- tritium and deuterium -- $^3$He, plasmas radiate more energy than fusion produces \cite[p.15]{AirNR}.

\subsection{Synchrotron radiation}
Overall, the field of synchrotron power loss within Tokamaks and Stellerators has not been studied extensively.  This phenomenon is not important for deuterium -- tritium reactors. Deuterium -- $^3$He reactors, where synchrotron radiation is significant, are still a project for the far future.  Thus, there has been little practical need to study synchrotron power loss.

In the discussion below, by \textbf{synchrotron radiation power} we mean the power of synchrotron radiation actually escaping plasma -- not the totality of emitted synchrotron power.  In most deuterium -- $^3$He reactors, only 1.5\% to 4.5\% of plasma synchrotron power actually escapes.  In Big ITER deuterium-tritium reactor, about 6\% to 8\% of synchrotron power escapes.  The rest is absorbed by plasma \cite[p.70]{BeegITER}.

Most plasma analysis in this section is done in zero dimensions.  In the \textbf{zero-dimensional analysis}, the plasma is examined as a homogenous medium, and all parameters are calculated per unit plasma volume\cite[p.4]{R3B4v01}.  Most of zero-dimensional analysis does not take reactor dimensions into consideration.  In case of synchrotron radiation, however, we must consider reactor major and minor radii.

An exact calculation of synchrotron radiation power per unit volume is beyond the scope of this work.  An approximate expression for synchrotron radiation power per unit volume is below \cite[p.4]{R3B4v01}.
    \be
    \label{1.03.06}
    \mathcal{P}_{_{\text{Synchrotron}}} \propto
    \sqrt{1-w_{_r}}\ B_{_T}^{2.5}\ T^{2.5}\ \sqrt{\frac{A\ n_e}{R}}
    \ \left[1+\sqrt{\frac{420\ keV}{A^2\ T}} \right],
    \ee
where $B_{_T}$ is the toroidal magnetic field within a Tokamak or Stellarator reactor, $A=R/a$ is the ratio of the major and major radii of reactor torus, and $w_{_r}$ is the wall reflectivity of synchrotron radiation \cite[p.1783]{ANST01}.

Substituting (\ref{1.02.01})  into (\ref{1.03.06}), we obtain
    \be
    \label{1.03.07}
    \begin{split}
    \mathcal{P}_{_{\text{Synchrotron}}} &\propto
    \sqrt{1-w_{_r}}\ B_{_T}^{2.5}\ T^{2.5}\ \sqrt{n_e}
    \ h(A,R,T) \propto
    \sqrt{1-w_{_r}}\ B_{_T}^{2.5}\ T^{2.5}\ \sqrt{\frac{P_{_g}}{T}}
    \ h(A,R,T) \\
    & \propto
    \sqrt{1-w_{_r}}\ B_{_T}^{2.5}\ T^{2.5}\ \sqrt{\frac{B_{_T}^2\ \beta}{T}}
    \ h(A,R,T) \propto
    \sqrt{1-w_{_r}}\ B_{_T}^{3.5}\ T^{2}\ \beta^{0.5}
    \ h(A,R,T).
    \end{split}
    \ee
where
    \be
    \label{1.03.08}
    \begin{split}
    h(A,R,T)=\sqrt{\frac{A}{R}}\ \left[1+\sqrt{\frac{420\ keV}{A^2\ T}}\right].
    \end{split}
    \ee

Recall the expression (\ref{1.02.19}) for fusion power per volume:
    \be
    \label{1.03.09}
    \mathcal{P} \propto
    x_{_r}\ \sigma_{_{P}}(T)\ B_{_T}^4 \beta^2.
    \ee
Dividing (\ref{1.03.08}) by (\ref{1.03.09}), we obtain the fraction of fusion power radiated away via synchrotron radiation.  This fraction is called \textbf{synchrotron power fraction}:
    \be
    \label{1.03.10}
    f_{_{\text{Synchrotron}}}=
    \frac{\mathcal{P}_{_{\text{Synchrotron}}}}{\mathcal{P}} \propto
    \frac{1}{x_{_r}}\ \sqrt{1-w_{_r}}\ B_{_T}^{-0.5}\ \beta^{-1.5}\ A^{0.5}\ R^{-0.5}
    \left[\frac{T^{2}}{\sigma_{_{P}}(T)}\ \left(1+\sqrt{\frac{420\ keV}{A^2\ T}}\right)\right].
    \ee

First, we calculate synchrotron radiation loss for deuterium -- $^3$He reactors.
In order to visualize the implications of (\ref{1.03.10}), we simplify that equation:
    \be
    \label{1.03.11}
    f_{_{\text{Synchrotron}}}^{\text{He}}\propto \frac{1}{x_{_r}}\
    \sqrt{1-w_{_r}}
    \ \left(\frac{B_{_T}}{1\ Tesla}\right)^{-0.5}
    \ \left(\frac{a}{1\ m}\right)^{-0.5}
    \ \beta^{-1.5}
    \ \mathcal{T}_{_{\text{He}}}(T,A),
    \ee
where
    \be
    \label{1.03.12}
    \mathcal{T}_{_{\text{He}}}(T,A)=
    \frac{
    \left[\frac{T^{2}}{\sigma_{_{P}}(T)}\ \left(1+\sqrt{\frac{420\ keV}{A^2\ T}}\right)\right]}
    {\left.
    \left[\frac{T^{2}}{\sigma_{_{P}}(T)}\ \left(1+\sqrt{\frac{420\ keV}{A^2\ T}}\right)\right]
    \right|_{A=1.4}^{T=50\ keV}
    }.
    \ee
The function $\mathcal{T}_{_{\text{He}}}(T,A)$ is tabulated in rows 5-7 of Table \ref{1.0T06} below.  Row 4 is the fraction of fusion power radiated away as Bremsstrahlung radiation.  This row is useful as the optimal choice of temperature optimizes between Bremsstrahlung and Synchrotron losses.
\begin{center}
  \begin{tabular}{|l|l|l|l|l|l|l|l|l|l|l|l|}
     \hline
     Quantity & Unit & & & & & & & & \\
     \hline
     T &$keV$
     & 30 & 40   &  50   & 60   & 70   &  80  & 90  & 100  \\
     \hline
     $\sigma_{_{P}}(T)$ &
     $10^{-2}\ bar^{-1}  s^{-1}$ &
     2.54 & 3.41 & 3.80  & 3.86 &  3.73 &  3.49 &  3.21 &  2.94 \\
     \hline
     $x_{_r}\ f_{_B}$ & &
     1.32 & 0.64 & 0.41  & 0.31 &  0.25 &  0.22 &  0.2  & 0.19  \\
     \hline
     $\mathcal{T}_{_{\text{He}}}(T,1.4)$ & &
     0.64 & 0.77 &  1.   & 1.33 &  1.79 &  2.39 &  3.17 &  4.15 \\
     $\mathcal{T}_{_{\text{He}}}(T,2)$ & &
     0.5  & 0.61 &  0.8  & 1.07 &  1.45 &  1.95 &  2.60 &  3.41 \\
     $\mathcal{T}_{_{\text{He}}}(T,3)$ & &
     0.39 &  0.48&   0.64& 0.87 &  1.18 &  1.60 &  2.15 &  2.84 \\
     \hline
   \end{tabular}
   \captionof{table}{Function $\mathcal{T}_{_{\text{He}}}(T,A)$} \label{1.0T06}
\end{center}

Second, we calculate synchrotron radiation loss for deuterium-tritium reactors.  Synchrotron radiation loss is insignificant in some deuterium -- tritium reactors.  As we can see from data in Tables \ref{1.0T01} and \ref{1.0T02}, the temperature term for deuterium -- tritium reactor is at least 680  times lower than the temperature term for deuterium -- $^3$He reactor.  Most works dealing with deuterium -- tritium fusion do not even mention synchrotron radiation.  Nevertheless, synchrotron radiation loss can be significant for deuterium -- tritium reactors with low $\beta$ and high operating temperature.  Moreover, given that 80\% of energy produced by deuterium -- tritium fusion is carried off by neutrons, even a small loss by synchrotron radiation can affect energy balance considerably.

Simplifying (\ref{1.03.10}), we obtain
    \be
    \label{1.03.13}
    f_{_{\text{Synchrotron}}}^{\text{T}}\propto \frac{1}{x_{_r}}\
    \sqrt{1-w_{_r}}
    \ \left(\frac{B_{_T}}{1\ Tesla}\right)^{-0.5}
    \ \left(\frac{a}{1\ m}\right)^{-0.5}
    \ \beta^{-1.5}
    \ \mathcal{T}_{_{\text{T}}}(T,A),
    \ee
where
    \be
    \label{1.03.14}
    \mathcal{T}_{_{\text{T}}}(T,A)=
    \frac{
    \left[\frac{T^{2}}{\sigma_{_{P}}(T)}\ \left(1+\sqrt{\frac{420\ keV}{A^2\ T}}\right)\right]}
    {\left.
    \left[\frac{T^{2}}{\sigma_{_{P}}(T)}\ \left(1+\sqrt{\frac{420\ keV}{A^2\ T}}\right)\right]
    \right|_{A=3}^{T=15\ keV}
    }.
    \ee
The function $\mathcal{T}_{_{\text{T}}}(T,A)$ is tabulated in rows 5-7 of Table \ref{1.0T07} below.
\begin{center}
  \begin{tabular}{|l|l|l|l|l|l|l|l|l|l|l|l|}
     \hline
     Quantity & Unit & & & & & & & & & \\
     \hline
     T &$keV$
     & 5    & 8    & 10   & 15   & 20   & 25  & 30  & 35  & 40 \\
     \hline
     $\sigma_{_{P}}(T)$ &
     $bar^{-1}  s^{-1}$ &
     1.43  & 2.53 &3.00  &3.24  &2.91  &2.46  &2.05  &1.68  & 1.37 \\
     \hline
     $x_{_r}\ f_{_B}$ &
     &0.19 &  0.052 &  0.032 &  0.016  & 0.012 &  0.010 &  0.009 &  0.009 &  0.009\\
     \hline
     $\mathcal{T}_{_{\text{T}}}(T,1.4)$ & &
     0.69 &  0.81 &  0.98 &  1.73  & 3.05 &  5.19 &  8.41 &  13.2 &  20.1\\
     $\mathcal{T}_{_{\text{T}}}(T,2)$ & &
     0.51 &  0.61 &  0.74 &  1.32  & 2.35 &  4.03 &  6.57 &  10.4 &  15.9\\
     $\mathcal{T}_{_{\text{T}}}(T,3)$ & &
     0.37 &  0.45 &  0.55 &  1.00  & 1.81 &  3.13 &  5.15 &  8.17 &  12.6\\
     \hline
   \end{tabular}
   \captionof{table}{Function $\mathcal{T}_{_{\text{T}}}(T,A)$} \label{1.0T07}
\end{center}
% T(2.9,12)=0.73

Maximum synchrotron power fractions for several proposed reactors are tabulated in Table \ref{1.0T08} below \cite[p.70]{BeegITER}.  The same source contains lower estimates for synchrotron power.  As we have mentioned, estimation of total synchrotron power loss for Tokamaks and Stellerators is still an unsolved problem.
\begin{center}
  \begin{tabular}{|l|l|l|l|l|l|l|l|l|l|l|}
  \hline
  Reactor   & Type     & $R$  & $A$  & $B_{_T}$ & T     & $\beta$  & $w_{_r}$ &
  $x_{_r}\ f_{_{\text{Synchrotron}}}$ \\
            &          & $m$  &      & $Tesla$  & $keV$ &          &          & \\
  \hline
  Big ITER  & D-T      & 8.14 & 2.91 & 5.7      & 12    & 0.030    &   0.7    & 0.013 \\
  ARIES III & D-$^3$He & 7.5  & 3.00 & 7.6      & 53    & 0.13     &   0.85   & 0.47  \\
  JOHNER 91 & D-$^3$He & 15   & 3.00 & 8.0      & 40    & 0.063    &   0.85   & 0.41  \\
  \hline
\end{tabular}
  \captionof{table}{Synchrotron power fractions of several proposed reactors} \label{1.0T08}
\end{center}

Based on Eq. (\ref{1.03.11}) and data presented in Table \ref{1.0T08}, we derive the following proportionality coefficient:
    \be
    \label{1.03.15}
    \begin{split}
    f_{_{\text{Synchrotron}}}^{\text{He}} &\lessapprox \frac{0.35}{x_{_r}}\
    \sqrt{1-w_{_r}}
    \ \left(\frac{B_{_T}}{1\ Tesla}\right)^{-0.5}
    \ \left(\frac{a}{1\ m}\right)^{-0.5}
    \ \beta^{-1.5}
    \ \mathcal{T}_{_{\text{He}}}(T,A)\\
    f_{_{\text{Synchrotron}}}^{\text{T}} &\approx \frac{6.7 \cdot 10^{-4}}{x_{_r}}\
    \sqrt{1-w_{_r}}
    \ \left(\frac{B_{_T}}{1\ Tesla}\right)^{-0.5}
    \ \left(\frac{a}{1\ m}\right)^{-0.5}
    \ \beta^{-1.5}
    \ \mathcal{T}_{_{\text{T}}}(T,A).
    \end{split}
    \ee
As we see from Eq. (\ref{1.03.15}) above, the proportionality constant for deuterium -- $^3$He fusion exceeds the proportionality constant for deuterium -- tritium fusion by a factor of 520.

In Table \ref{1.0T09} below, we present synchrotron power fraction for Big ITER reactor as a function of plasma temperature.  In row 6, $f_{_{\gamma}}$ is the total loss due to photon radiation, which is the sum of rows 4 and 5.
\begin{center}
  \begin{tabular}{|l|l|l|l|l|l|l|l|l|l|l|l|}
     \hline
     Quantity & Unit & & & & & & & & & \\
     \hline
     T &$keV$
     & 5    & 8    & 10   & 15   & 20   & 25  & 30  & 35  & 40 \\
     \hline
     $\sigma_{_{P}}(T)$ &
     $bar^{-1}  s^{-1}$ &
     1.43  & 2.53 &3.00  &3.24  &2.91  &2.46  &2.05  &1.68  & 1.37 \\
     \hline
     $x_{_r}\ f_{_B}$ &
     &0.19 &  0.052 &  0.032 &  0.016  & 0.012 &  0.010 &  0.009 &  0.009 &  0.009\\
     \hline
     $x_{_r}\ f_{_{\text{Synchrotron}}}$ & &
     0.005 &  0.006 &  0.007 &  0.013 &  0.024 &  0.041 &  0.068 &  0.108 &  0.167\\
     \hline
     $x_{_r}\ f_{_{\gamma}}$ &
     &0.20 &  0.058 &  0.039 &  0.029  & 0.036 &  0.051 &  0.077 &  0.117 & 0.176 \\
     \hline
   \end{tabular}
   \captionof{table}{Radiation losses for Big ITER} \label{1.0T09}
\end{center}
As we see from the table above, synchrotron radiation loss does hinder reactor operation at high temperatures.

\section{Transport energy loss}

\subsection{Fusion power gain $Q$}

\textbf{Fusion power gain} denoted by $Q$ is defined as the ratio of thermal energy produced by fusion to the heating power supplied to the plasma \cite[p.14]{Tokamak}.  A nuclear power station must have fusion power gain of at least 3.  In order to calculate $Q$ we must calculate the power generated by nuclear fusion as well as external power needed to heat the plasma.  External plasma heating power per unit volume is
      \be
      \label{1.04.01}
      \mathcal{P}_{_{\text{heat}}}=\mathcal{P}_{_{\text{loss}}}-\mathcal{P}_{_{\text{fh}}}.
      \ee
In Eq. (\ref{1.04.01}) above, $\mathcal{P}_{_{\text{loss}}}$ is the rate of plasma energy loss per unit volume.  $\mathcal{P}_{_{\text{fh}}}$ is the fusion power which is converted into plasma thermal energy.  Notice, that if the expression in Eq. (\ref{1.04.01}) is negative, then it means that plasma temperature is rising.

In order to understand how plasma loses energy by conduction, we must understand plasma behavior within the reactor.  Magnetic confinement can not be perfect, thus the plasma fills up the volume of a Tokamak or Stellarator torus and physically interacts with the walls \cite[p.81]{FusBk1}.  Thermal energy is transported within plasma by conduction, convection, and turbulent flow \cite[p.81]{FusBk1}.  The outer layer of plasma conducts thermal energy to the walls.

Even though the plasma temperature is hundreds of millions $^o$K, the plasma neither melts nor vaporizes the container walls.  Some solid surfaces can be in contact with very hot, yet rarefied gases without overheating or sustaining damage.  A common experiment demonstrating this effect is boiling water in a paper cup.  The cup does not get charred as the heat from the fire is immediately conducted into water.  Another experiment from a different branch of science is the descent of a space capsule into the atmosphere.  Even though such a capsule may encounter a stream of gas at a temperature of several thousand degrees Kelvin, the rarefied air does not damage the capsule.

Energy loss of plasma per unit volume by conduction is \cite[p.17]{Tokamak}
      \be
      \label{1.04.02}
      \mathcal{P}_{_{\text{loss}}}=\frac{\mathcal{W}}{\tau_{_E}}=\frac{1.5\ P_{_g}}{\tau_{_E}},
      \ee
where $\tau_{_E}$ is called \textbf{energy confinement time}, and $\mathcal{W}$ is kinetic energy density of plasma.  For a fully ionized plasma, $\mathcal{W}=1.5\ P_{_g}$.  The fusion power per unit volume used to heat plasma is derived from (\ref{1.02.16}):
      \be
      \label{1.04.03}
      \mathcal{P}_{_{\text{fh}}}=
      f_{_{\text{heat}}}\ \mathcal{P}=
      \frac{x_{_r}}{4}\ f_{_{\text{heat}}}\ \sigma_{_{P}}(T)\ P_{_g}^2,
      \ee
whaere $f_{_{\text{heat}}}$ is the fraction of fusion power which is not lost to neutron,  Bremsstrahlung and synchrotron radiation.  It is given by
      \be
      \label{1.04.04}
      f_{_{\text{heat}}}=1-f_{_{\text{n}}}-f_{_{\gamma}},
      \ee
where $f_{_{\text{n}}}$ is the neutronicity or the fraction of fusion power carried away by neutrons, and $f_{_{\gamma}}$ is the fraction of fusion power radiated away via synchrotron and Bremsstrahlung radiation.

The vales of $f_{_{\text{heat}}}$ depend on plasma temperature and composition.  For deuterium -- tritium fusion, neutronicity is 80\%.  The value of $f_{_{\gamma}}$ is dependent on reactor dimensions, $\beta$, and operating temperature.
For Big ITER reactor, which would operate at $T=1.3\ keV$, $f_{_B}=0.067$, $f_{_{\text{Synchrotron}}}=0.013$, and $f_{_{\gamma}}=0.079$ \cite{BigITER,BeegITER}.
Based on the values of neutronicity and synchrotron and Bremsstrahlung power fractions presented above, we conclude that for Big ITER reactor
      \be
      \label{1.04.05}
      f_{_{\text{heat}}}=0.12.
      \ee
As we show in Subsection {\rr 6.2} below, deuterium-$^3$He reactors may have higher $f_{_{\text{heat}}} \le 0.45$.

Combining (\ref{1.02.16}) and (\ref{1.04.03}), we obtain the fusion power gain:
      \be
      \label{1.04.06}
      \begin{split}
      Q&=\frac{\mathcal{P}}{\mathcal{P}_{_{\text{heat}}}}
       =\frac{\mathcal{P}}{\mathcal{P}_{_{\text{loss}}}-\mathcal{P}_{_{\text{fh}}}}
       =\frac{\sigma_{_{P}}(T)\ P_{_g}^2}{\frac{1.5\ P_{_g}}
       {\tau_{_E}}-\frac{1}{4}\ f_{_{\text{heat}}}\ \sigma_{_{P}}(T)\ P_{_g}^2}
       =
       \frac{P_{_g}\ \tau_{_E}\ f_{_{\text{heat}}}^{-1}}
       {1.5\ f_{_{\text{heat}}}^{-1}\ \big(\sigma_{_{P}}(T)\big)^{-1}
       -\frac{1}{4}\ P_{_g}\ \tau_{_E}}.
        \end{split}
        \ee
In order to understand the physical meaning of Eq. (\ref{1.04.06}), we introduce the following concepts.  The \textbf{Lawson pressure criterion} is
        \be
        \label{1.04.07}
        C_{_{LP}}=P_{_{g}} \ \tau_{_E}.
        \ee
The \textbf{Lawson pressure ignition criterion} is
        \be
        \label{1.04.08}
        C_{_{LPI}}=\frac{6}{f_{_{\text{heat}}}\ \sigma_{_{P}}(T)}.
        \ee
Notice, that Lawson pressure ignition criterion increases as more fuel gets consumed.  As we have shown in Subsection {\rr 2.4}, as more fuel is consumed, $x_{_r}$ decreases.  As we have shown in Section {\rr 3}, decreasing $x_{_r}$ causes the fraction of fusion power radiated away via synchrotron and Bremsstrahlung radiation to increase.  This causes $f_{_{\text{heat}}}$ to decrease and $C_{_{LPI}}$ to increase.

Substituting (\ref{1.04.05}) into (\ref{1.04.08}) above, we find that for a deuterium -- tritium reactor such as Big ITER $C_{_{LPI}}=16\ bar \cdot s$.  Given that $C_{_{LPI}}$ varies over the course of fuel burn cycle, $16\ bar \cdot s$ is the time average.

Substituting (\ref{1.04.07}) and (\ref{1.04.08}) into (\ref{1.04.06}), we obtain
        \be
        \label{1.04.09}
        \left\{
        \begin{split}
        Q&=\frac{C_{_{LP}}\ f_{_{\text{heat}}}^{-1}}
        {C_{_{LPI}}-C_{_{LP}}}
        \qquad \text{if} \qquad C_{_{LPI}} > C_{_{LP}}
        \\
        Q&=\infty
        \hskip2.3cm
        \text{if} \qquad C_{_{LPI}} \le C_{_{LP}}
        \end{split}
        \right.
        \ee
If the reactor Lawson pressure criterion is the same or higher than the Lawson Pressure Ignition Criterion, then the fusion is ignited -- the reaction continues without an external heat source.

Most works on nuclear fusion define the Lawson criterion based on particle density \cite[p.10]{FusBk1}:
    \be
    \label{1.04.10}
    C_{_{Ld}}=n\ T\ \tau_{_E},
    \ee
where  $n$ is the ion number density.  The Lawson Criterion for working deuterium -- tritium reactors has to be $C_{_{Ld}} \ge 3 \cdot 10^{21}\ \ keV\ s\ m^{-3}$ \cite[p.3]{TSurvey03}.  Using (\ref{1.02.01}) and (\ref{1.04.10}), we express Lawson pressure criterion in terms of Lawson density criterion:
    \be
    \label{1.04.11}
    C_{_{LP}}=P_{_{g}} \ \tau_{_E}=(1+Z)\ R_{_{g}}\ n\ \ T \tau_{_E}=(1+\overline{Z})\ R_{_{g}}\ C_{_{Ld}}.
    \ee

\subsection{Confinement time and Lawson criterion scaling}
Plasma held by magnetic fields inside a Tokamak or Stellarator is a fluid governed by very complex physics.  Thus, energy confinement time is dependent on many factors, including plasma heating mechanism.  Plasma heated only by electric current is defined to be in \textbf{ohmic mode} \cite[p.55]{FusBk1}.  Ohmic heating alone can not raise plasma temperature beyond about 2 $keV$ \cite[p.64]{FusBk1}.

As we have mentioned in Subsection {\rr 3.1}, plasma can be heated by neutral particle beams or microwaves.  This mode of heating produces plasma in \textbf{L-mode confinement} \cite[p.55]{FusBk1}.  In L-mode, L denotes "low".  It has lower energy confinement time than ohmic mode \cite[p.20]{T5}.

Ohmic and L-mode confinement have been known since the 1960s.  Then a new mode was discovered \cite[p.55]{FusBk1}:
    \begin{quote}
    It was found that when sufficient power was applied to an L-mode discharge that the
    discharge made an abrupt transition in which the edge transport was apparently
    reduced, leading to edge pedestals in the temperature and density. The effect of this
    was to produce roughly a doubling of the confinement time.
    \end{quote}
The new mode of confinement is called \textbf{H-mode}.
Other modes with confinement times longer than H-mode are possible.  These modes are unstable and thus non-usable \cite[p.58]{T5}.
All proposed reactors would use H-mode confinement.  In the rest of the work, we consider only this mode.

A general model for confinement time is given by \cite[p.58]{T5}
    \be
    \label{1.04.12}
    \tau_{_{E}} \propto
    I^{\alpha_{_I}}\
    B_{_T}^{\alpha_{_B}}\
    P^{\alpha_{_P}}\
    n^{\alpha_{_n}}\
    R^{\alpha_{_R}}\
    \overline{M}^{\alpha_{_M}}\
    A^{\alpha_{_A}}\
    \kappa^{\alpha_{_\kappa}},
    \ee
where
$P$ is the plasma heating power,
$\overline{M}$ is the mean plasma ion mass in $amu$,
$A=a/R$ is the aspect ratio.
From (\ref{1.02.03}), we obtain
    \be
    \label{1.04.13}
    n \propto \frac{P_{_{g}}}
    {(1+\overline{Z})\ T}.
    \ee
At this point, we express the plasma gas pressure in terms of magnetic field.  Using (\ref{1.02.04}) and (\ref{1.02.05}), we obtain
    \be
    \label{1.04.14}
    P_{_g} \propto B_{_T}^2 \beta.
    \ee
Substituting (\ref{1.04.14}) into (\ref{1.04.13}), we get
    \be
    \label{1.04.15}
    n \propto \frac{B_{_T}^2 \beta}
    {(1+\overline{Z})\ T}.
    \ee

The \textbf{safety factor} is defined as \cite[p. 268]{Tokamak}
    \be
    \label{1.04.16}
    q=\frac{S_{_F}}{A}\ \left(\frac{I}{10^6\ A}\right)^{-1}
    \left(\frac{R}{1\ m}\right) \left(\frac{B_{_T}}{1\ Tesla}\right),
    \ \ \text{hence} \ \
    I \propto \big(S_{_F}\ A\big) \frac{R\ B_{_T}}{A^2\ q}=
    \hat{S}_{_F}\ \frac{R\ B_{_T}}{A^2\ q}.
    \ee
In Eq. (\ref{1.04.16}) above, $S_{_F}$ is the \textbf{shape factor}.  It is approximated by \cite[p. 268]{Tokamak}:
    \be
    \label{1.04.17}
    S_{_F} \approx
    \frac{2.5}{A}\
    \Big(1+\kappa^2 \left(1+2\ \delta^2-1.2\ \delta^3 \right) \Big)\
    \frac{1.17-0.65/A}{\Big(1-1/A^2 \Big)^2},
    \ee
where $\delta$ is triangularity of plasma.  The function $f(A)$ is
    \be
    \label{1.04.18}
    f(A)=\frac{1.17-0.65/A}{\Big(1-1/A^2 \Big)^2}.
    \ee
Another estimate for $f(A)$ is \cite[p.1226]{ST30}
    \be
    \label{1.04.19}
    f_{_1}(A)=1.17 \sqrt{\frac{A}{A-1}}.
    \ee
Projects for several spherical Tokamaks listed in \cite[p. 14]{ST19} agree with  Eq. (\ref{1.04.17}) to within a maximal error of 23\% and an average error of 13\%.

We also introduce scaled shape factor
    \be
    \label{1.04.20}
    \hat{S}_{_F}=S_{_F}\ A.
    \ee

Substituting (\ref{1.04.15}) and (\ref{1.04.16}) into (\ref{1.04.12}), we obtain
    \be
    \label{1.04.21}
    \tau_{_{E}} \propto
    \hat{S}_{_F}^{\alpha_{_I}}\
    P^{\alpha_{_P}}\
    B_{_T}^{\big[\alpha_{_B}+2 \alpha_{_n} + \alpha_{_I}\big]}\
    R^{^{\big[\alpha_{_R}+\alpha_{_I} \big]}}\
    A^{^{\big[\alpha_{_A}-2\alpha_{_I} \big]}}\
    \beta^{^{\alpha_{_n}}}\
    q^{-\alpha_{_I}}\
    T^{-\alpha_{_n}}\
    \kappa^{\alpha_{_\kappa}}\
    \overline{M}^{\alpha_{_M}}\
    (\overline{Z}+1)^{-\alpha_{_n}}\
    \ee

For a steady-state reactor, conductive power loss is \cite[p.82]{AF01}:
    \be
    \label{1.04.22}
    P=\frac{W}{\tau_{_E}},
    \ee
where $W$ is the total thermal energy of plasma.  It is given by
    \be
    \label{1.04.23}
    W=1.5\ P_{_g} V_{_g},
    \ee
where $V_{_g}$ is the plasma volume given by
    \be
    \label{1.04.24}
    V_{_g}=2\ \pi^2\ R\ a^2\ \kappa \propto R^3\ A^{-2}\ \kappa
    \ee
Substituting (\ref{1.04.14}), (\ref{1.04.23}), and (\ref{1.04.24}) into (\ref{1.04.22}), we obtain
    \be
    \label{1.04.25}
    P \propto \tau_{_E}^{-1}\
    \Big( B_{_T}^2\ \beta \Big)\
    \Big( R^3\ A^{-2}\ \kappa \Big)=
    \tau_{_E}^{-1}\ B_{_T}^2\ R^3\ A^{-2}\ \beta\ \kappa.
    \ee

Substituting (\ref{1.04.25}) into (\ref{1.04.21}), we obtain
    \be
    \label{1.04.26}
    \begin{split}
    \tau_{_E} \propto&
    \tau_{_E}^{-\alpha_{_P}}\
    \hat{S}_{_F}^{\alpha_{_I}}\
    B_{_T}^{\big[2\alpha_{_P}+\alpha_{_B}+2 \alpha_{_n} + \alpha_{_I}\big]}\
    R^{^{\big[3\alpha_{_P}+\alpha_{_R}+\alpha_{_I} \big]}}\
    A^{^{\big[-2\alpha_{_P}+\alpha_{_A}-2\alpha_{_I} \big]}}\\
    &\beta^{^{\alpha_{_P}+\alpha_{_n}}}\
    q^{-\alpha_{_I}}\
    T^{-\alpha_{_n}}\
    \kappa^{^{\big[\alpha_{_\kappa}+\alpha_{_P}\big]}}\
    \overline{M}^{\alpha_{_M}}\
    (\overline{Z}+1)^{-\alpha_{_n}}.
    \end{split}
    \ee
Solving (\ref{1.04.26}), we obtain
    \be
    \label{1.04.27}
    \begin{split}
    \tau_{_E} \propto&
    \hat{S}_{_F}^{\big[\frac{\alpha_{_I}}{1+\alpha_{_P}}\big]}\
    B_{_T}^{\big[\frac{2\alpha_{_P}+\alpha_{_B}+2 \alpha_{_n} + \alpha_{_I}}
    {1+\alpha_{_P}}\big]}\
    R^{^{\big[\frac{3\alpha_{_P}+\alpha_{_R}+\alpha_{_I}}
    {1+\alpha_{_P}} \big]}}\
    A^{^{\big[\frac{-2\alpha_{_P}+\alpha_{_A}-2\alpha_{_I}}
    {1+\alpha_{_P}} \big]}}\
    \beta^{^{\big[\frac{\alpha_{_P}+\alpha_{_n}}{1+\alpha_{_P}}\big]}}\
    \\
    &
    q^{^{\big[\frac{-\alpha_{_I}}{1+\alpha_{_P}}\big]}}\
    T^{^{\big[\frac{-\alpha_{_n}}{1+\alpha_{_P}}\big]}}\
    \kappa^{^{\big[\frac{\alpha_{_\kappa}+\alpha_{_P}}{1+\alpha_{_P}}\big]}}\
    \overline{M}^{^{\big[\frac{\alpha_{_M}}{1+\alpha_{_P}}\big]}}\
    (\overline{Z}+1)^{^{\big[\frac{-\alpha_{_n}}{1+\alpha_{_P}}\big]}}.
    \end{split}
    \ee
Given the complexity of (\ref{1.04.27}), we express it in a logarithmic form
    \be
    \label{1.04.28}
    \begin{split}
    \ln \tau_{_E} &\propto
    \left[\frac{\alpha_{_I}}{1+\alpha_{_P}}\right] \ln \hat{S}_{_F}+
    \left[\frac{2\alpha_{_P}+\alpha_{_B}+2 \alpha_{_n} + \alpha_{_I}}
    {1+\alpha_{_P}}\right] \ln B_{_T}+
    \left[\frac{3\alpha_{_P}+\alpha_{_R}+\alpha_{_I}}
    {1+\alpha_{_P}} \right] \ln R\\
    &+\left[\frac{-2\alpha_{_P}+\alpha_{_A}-2\alpha_{_I}}
    {1+\alpha_{_P}} \right] \ln A+
    \left[\frac{\alpha_{_P}+\alpha_{_n}}{1+\alpha_{_P}}\right] \ln \beta+
    \left[\frac{-\alpha_{_I}}{1+\alpha_{_P}}\right] \ln q\\
    &+\left[\frac{\alpha_{_\kappa}+\alpha_{_P}}{1+\alpha_{_P}}\right]
    \ln \kappa+
    \left[\frac{-\alpha_{_n}}{1+\alpha_{_P}}\right] \ln T+
    \left[\frac{\alpha_{_M}}{1+\alpha_{_P}} \right] \ln \overline{M}+
    \left[\frac{-\alpha_{_n}}{1+\alpha_{_P}}\right] \ln (\overline{Z}+1).
    \end{split}
    \ee
Substituting (\ref{1.04.11}) and (\ref{1.04.14}) into (\ref{1.04.28}) for Lawson pressure criterion, we obtain
    \be
    \label{1.04.29}
    \begin{split}
    \ln C_{_{LP}} &\propto
    \left[\frac{\alpha_{_I}}{1+\alpha_{_P}}\right] \ln \hat{S}_{_F}+
    \left[2+\frac{2\alpha_{_P}+\alpha_{_B}+2 \alpha_{_n} + \alpha_{_I}}
    {1+\alpha_{_P}}\right] \ln B_{_T}+
    \left[\frac{3\alpha_{_P}+\alpha_{_R}+\alpha_{_I}}
    {1+\alpha_{_P}} \right] \ln R\\
    &+\left[\frac{-2\alpha_{_P}+\alpha_{_A}-2\alpha_{_I}}
    {1+\alpha_{_P}} \right] \ln A+
    \left[\frac{1+2\alpha_{_P}+\alpha_{_n}}{1+\alpha_{_P}}\right] \ln \beta+
    \left[\frac{-\alpha_{_I}}{1+\alpha_{_P}}\right] \ln q\\
    &+\left[\frac{\alpha_{_\kappa}+\alpha_{_P}}{1+\alpha_{_P}}\right]
    \ln \kappa+
    \left[\frac{-\alpha_{_n}}{1+\alpha_{_P}}\right] \ln T+
    \left[\frac{\alpha_{_M}}{1+\alpha_{_P}} \right] \ln \overline{M}+
    \left[\frac{-\alpha_{_n}}{1+\alpha_{_P}}\right] \ln (\overline{Z}+1).
    \end{split}
    \ee
Substituting (\ref{1.04.28}) into (\ref{1.04.25}), we obtain the heating power
    \be
    \label{1.04.30}
    \begin{split}
    \ln P &\propto
    \left[\frac{-\alpha_{_I}}{1+\alpha_{_P}}\right] \ln \hat{S}_{_F}+
    \left[2-\frac{2\alpha_{_P}+\alpha_{_B}+2 \alpha_{_n} + \alpha_{_I}}
    {1+\alpha_{_P}}\right] \ln B_{_T}+
    \left[3-\frac{3\alpha_{_P}+\alpha_{_R}+\alpha_{_I}}
    {1+\alpha_{_P}} \right] \ln R\\
    &+\left[-2+\frac{2\alpha_{_P}-\alpha_{_A}+2\alpha_{_I}}
    {1+\alpha_{_P}} \right] \ln A+
    \left[1-\frac{\alpha_{_P}+\alpha_{_n}}{1+\alpha_{_P}}\right] \ln \beta+
    \left[\frac{\alpha_{_I}}{1+\alpha_{_P}}\right] \ln q\\
    &+\left[1-\frac{\alpha_{_\kappa}+\alpha_{_P}}{1+\alpha_{_P}}\right]
    \ln \kappa+
    \left[\frac{\alpha_{_n}}{1+\alpha_{_P}}\right] \ln T+
    \left[\frac{-\alpha_{_M}}{1+\alpha_{_P}} \right] \ln \overline{M}+
    \left[\frac{\alpha_{_n}}{1+\alpha_{_P}}\right] \ln (\overline{Z}+1).
    \end{split}
    \ee
Simplifying (\ref{1.04.30}), we obtain
    \be
    \label{1.04.31}
    \begin{split}
    \ln P &\propto
    \left[\frac{-\alpha_{_I}}{1+\alpha_{_P}}\right] \ln \hat{S}_{_F}+
    \left[\frac{2-\alpha_{_B}-2 \alpha_{_n} - \alpha_{_I}}
    {1+\alpha_{_P}}\right] \ln B_{_T}+
    \left[\frac{3-\alpha_{_R}-\alpha_{_I}}
    {1+\alpha_{_P}} \right] \ln R\\
    &+\left[\frac{-2-\alpha_{_A}+2\alpha_{_I}}
    {1+\alpha_{_P}} \right] \ln A+
    \left[\frac{1-\alpha_{_n}}{1+\alpha_{_P}}\right] \ln \beta+
    \left[\frac{\alpha_{_I}}{1+\alpha_{_P}}\right] \ln q\\
    &+\left[\frac{1-\alpha_{_\kappa}}{1+\alpha_{_P}}\right]
    \ln \kappa+
    \left[\frac{\alpha_{_n}}{1+\alpha_{_P}}\right] \ln T+
    \left[\frac{-\alpha_{_M}}{1+\alpha_{_P}} \right] \ln \overline{M}+
    \left[\frac{\alpha_{_n}}{1+\alpha_{_P}}\right] \ln (\overline{Z}+1).
    \end{split}
    \ee

Below, we express conductive power loss and Lawson pressure criterion in terms of normalized $\beta$, which is defined as \cite[p.9]{FusBk3}
    \be
    \label{1.04.32}
    \beta_{_N}=\frac{100\ \beta}{A}\ \left(\frac{I}{10^6\ A}\right)^{-1}
    \left(\frac{R}{1\ m}\right) \left(\frac{B_{_T}}{1\ Tesla}\right).\\
    \ee
Substituting (\ref{1.04.16}) into (\ref{1.04.32}) we obtain
    \be
    \label{1.04.33}
    \begin{split}
    \beta_{_N}=\frac{100\ q\ \beta}{S_{_F}}=\frac{100\ q\ \beta\ A}{\hat{S}_{_F}},
    \qquad \text{hence} \qquad
    \beta \propto \frac{\beta_{_N}\ S_{_F}}{q\ A}.
    \end{split}
    \ee
Both $\beta$ and $\beta_{_N}$ are dimensionless parameters.
Substituting (\ref{1.04.33}) into (\ref{1.04.29}), we obtain
    \be
    \label{1.04.34}
    \begin{split}
    \ln C_{_{LP}} &\propto
    \left[\frac{1+2\alpha_{_P}+\alpha_{_n}+\alpha_{_I}}{1+\alpha_{_P}}\right]
    \ln \hat{S}_{_F}+
    \left[2+\frac{2\alpha_{_P}+\alpha_{_B}+2 \alpha_{_n} + \alpha_{_I}}
    {1+\alpha_{_P}}\right] \ln B_{_T}\\
    &+
    \left[\frac{3\alpha_{_P}+\alpha_{_R}+\alpha_{_I}}
    {1+\alpha_{_P}} \right] \ln R
    +\left[\frac{-1-4\alpha_{_P}+\alpha_{_A}-2\alpha_{_I}-\alpha_{_n}}
    {1+\alpha_{_P}} \right] \ln A\\
    &+\left[\frac{1+2\alpha_{_P}+\alpha_{_n}}{1+\alpha_{_P}}\right] \ln \beta_{_N}+
    \left[\frac{-1-2\alpha_{_P}-\alpha_{_n}-\alpha_{_I}}{1+\alpha_{_P}}\right]
    \ln q\\
    &+\left[\frac{\alpha_{_\kappa}+\alpha_{_P}}{1+\alpha_{_P}}\right]
    \ln \kappa+
    \left[\frac{-\alpha_{_n}}{1+\alpha_{_P}}\right] \ln T+
    \left[\frac{\alpha_{_M}}{1+\alpha_{_P}} \right] \ln \overline{M}+
    \left[\frac{-\alpha_{_n}}{1+\alpha_{_P}}\right] \ln (\overline{Z}+1).
    \end{split}
    \ee
Substituting (\ref{1.04.33}) into (\ref{1.04.31}), we obtain
    \be
    \label{1.04.35}
    \begin{split}
    \ln P &\propto
    \left[\frac{1-\alpha_{_n}-\alpha_{_I}}{1+\alpha_{_P}}\right] \ln \hat{S}_{_F}+
    \left[\frac{2-\alpha_{_B}-2 \alpha_{_n} - \alpha_{_I}}
    {1+\alpha_{_P}}\right] \ln B_{_T}+
    \left[\frac{3-\alpha_{_R}-\alpha_{_I}}
    {1+\alpha_{_P}} \right] \ln R\\
    &+\left[\frac{-3+\alpha_{_n}-\alpha_{_A}+2\alpha_{_I}}
    {1+\alpha_{_P}} \right] \ln A+
    \left[\frac{1-\alpha_{_n}}{1+\alpha_{_P}}\right] \ln \beta+
    \left[\frac{-1+\alpha_{_n}+\alpha_{_I}}{1+\alpha_{_P}}\right] \ln q\\
    &+\left[\frac{1-\alpha_{_\kappa}}{1+\alpha_{_P}}\right]
    \ln \kappa+
    \left[\frac{\alpha_{_n}}{1+\alpha_{_P}}\right] \ln T+
    \left[\frac{-\alpha_{_M}}{1+\alpha_{_P}} \right] \ln \overline{M}+
    \left[\frac{\alpha_{_n}}{1+\alpha_{_P}}\right] \ln (\overline{Z}+1).
    \end{split}
    \ee

One model for Tokamak energy confinement time in good agreement with experimental data is called ``IPB98(y,2)".  It is given by \cite[p.212]{FusBk1}
% H-mode
    \be
    \label{1.04.36}
    \tau_{_{ET}} \propto
    I^{0.93}\
    B_{_T}^{0.15}\
    P^{-0.69}\
    n^{0.41}\
    R^{1.97}\
    \overline{M}^{0.19}\
    A^{-0.58}\
    \kappa^{0.78}.
    \ee
Substituting (\ref{1.04.36}) into (\ref{1.04.30}) and (\ref{1.04.29}), we obtain scaling laws for power and Lawson pressure criterion based on ``IPB98(y,2)":
    \be
    \label{1.04.37}
    \begin{split}
    P_{_{\text{Tokamak}}} \propto &\hat{S}_{_F}^{-1.10} B_{_T}^{0.32} R^{0.32} A^{-0.48} \beta_{_N}^{1.90} q^{1.10} \kappa^{0.71} T^{1.32} \overline{M}^{-0.61} (\overline{Z}+1)^{1.32}\\
    C_{_{LP}} \propto &\hat{S}_{_F}^{3.10} B_{_T}^{3.68} R^{2.68} A^{-3.52} \beta_{_N}^{0.10} q^{-3.1} \kappa^{0.29} T^{-1.32} \overline{M}^{0.61} (\overline{Z}+1)^{-1.32}.
    \end{split}
    \ee
Notice, that $P_{_{\text{Tokamak}}}$ is the conductive power flux to Tokamak walls.  This quantity is not directly related to the fusion power within Tokamak.

A more recent scaling model is \cite[p. 131]{Tokamak}
% H-mode
    \be
    \label{1.04.38}
    \tau_{_{ET}} \propto
    I^{0.86}\
    B_{_T}^{0.21}\
    P^{-0.65}\
    n^{0.41}\
    R^{1.99}\
    \overline{M}^{0.08}\
    A^{-0.68}
    \kappa^{0.84}.
    \ee
Substituting (\ref{1.04.38}) into (\ref{1.04.30}) and (\ref{1.04.29}), we obtain scaling laws for power and Lawson pressure criterion based on the later model
    \be
    \label{1.04.39}
    \begin{split}
    P_{_{\text{Tokamak}}} \propto &\hat{S}_{_F}^{-0.71} B_{_T}^{0.43} R^{0.43} A^{-0.60} \beta_{_N}^{1.74} q^{0.71} \kappa^{0.46} T^{1.11} \overline{M}^{-0.23} (\overline{Z}+1)^{1.11}\\
    C_{_{LP}} \propto &\hat{S}_{_F}^{2.71} B_{_T}^{3.57} R^{2.57} A^{-3.4} \beta_{_N}^{0.26} q^{-2.71} \kappa^{0.54} T^{-1.11} \overline{M}^{0.23} (\overline{Z}+1)^{-1.11}.
    \end{split}
    \ee

A theoretical model is \cite[p. A1]{ST13}
    \be
    \label{1.04.40}
    \tau_{_{ET}} \propto
    I^{1.00}\
    B_{_T}^{0.00}\
    P^{-0.50}\
    n^{0.50}\
    R^{2.00}\
    \overline{M}^{0}\
    A^{-0.50}\
    \kappa^{0.75}.
    \ee
Substituting (\ref{1.04.40}) into (\ref{1.04.30}) and (\ref{1.04.29}), we obtain scaling laws for power and Lawson pressure criterion based on the later model
    \be
    \label{1.04.41}
    \begin{split}
    P_{_{\text{Tokamak}}} \propto &\hat{S}_{_F}^{-1} B_{_T}^{  0} R^{  0} A^{  0} \beta_{_N}^{  1} q^{  1} \kappa^{0.5} T^{  1} \overline{M}^{ -0} (\overline{Z}+1)^{  1}\ \ \ \ \
    =\hat{S}_{_F}^{-1} \beta_{_N}\ q\ \kappa^{0.5}\ T\ (\overline{Z}+1)
    \\
    C_{_{LP}} \propto &\hat{S}_{_F}^{3} B_{_T}^{  4} R^{  3} A^{ -4} \beta^{  1} q^{ -3} \kappa^{0.5} T^{ -1} \overline{M}^{  0} (\overline{Z}+1)^{ -1}
    =\hat{S}_{_F}^{3} B_{_T}^{  4} R^{  3} A^{ -4} \beta_{_N}\ q^{ -3} \kappa^{0.5} T^{ -1}
    (\overline{Z}+1)^{ -1}.
    \end{split}
    \ee

One model for Stellarator energy confinement time in good agreement with experimental data is called ``ISS04v3".
It is given by \cite{StelConf}
    \be
    \label{1.04.42}
    \begin{split}
    \tau_{_{ES}} \propto
    B_{_T}^{0.85}\
    P^{-0.61}\
    n^{0.55}
    R^{2.97}\
    \iota^{0.41}\
    A^{-2.33},
    \end{split}
    \ee
where
$\iota$ -- \textbf{magnetic helicity} or rotational transform of the field lines and  $q=1/\iota$ is the safety factor \cite[p.2]{Tok2}.
Substituting (\ref{1.04.16}) into (\ref{1.04.42}), we obtain ``ISS04v3" in terms of the model (\ref{1.04.12}):
    \be
    \label{1.04.43}
    \tau_{_{ES}} \propto
    I^{0.41}\
    B_{_T}^{0.44}\
    P^{-0.61}\
    n^{0.55}\
    R^{2.56}\
    \overline{M}^{0.00}\
    A^{-1.51}.
    \ee
Substituting (\ref{1.04.43}) into (\ref{1.04.30}) and (\ref{1.04.29}), we obtain scaling laws for power and Lawson pressure criterion based on ``ISS04v3":
    \be
    \label{1.04.44}
    \begin{split}
    P_{_{\text{Stellarator}}} \propto &B_{_T}^{0.13} R^{0.08} A^{0.85} \beta^{1.15} q^{1.05} T^{1.41} (\overline{Z}+1)^{1.41}\\
    C_{_{LP}} \propto &B_{_T}^{3.87} R^{2.92} A^{-2.85} \beta^{0.85} q^{-1.05} T^{-1.41} (\overline{Z}+1)^{-1.41}.
    \end{split}
    \ee
The term $\kappa$ is ignored, since $\kappa=1$ for Stellerator reactors.  The shape factor is also irrelevant to Stellarators.  Moreover, the result is expressed in terms of $\beta$, as normalized $\beta$ is irrelevant to Stellarators.
Notice, that $P_{_{\text{Stellarator}}}$ is the conductive power flux to Stellarator walls.  This quantity is not directly related to the fusion power within Stellarator.

Other sources make predictions of Tokamak and Stellarator fusion power based on \textbf{zero-dimensional model} -- analysis of fusion power balance per unit volume. \cite{R3B4v01}.  The total fusion power of a Tokamak is proportional to \cite[p.3]{R3B4v02}:
    \be
    \label{1.04.45}
    P_{_{_{\text{fusion}}}} \propto \frac{\beta_{_N}^2\ B_{_T}^4\ R^3}{q^2\ A^4}.
    \ee
The Lawson pressure criterion of a Tokamak is proportional to \cite[p.3]{R3B4v02}:
    \be
    \label{1.04.46}
    C_{_{LP}} \propto \frac{H^{3.23}\ \beta_{_N}^{0.1}\ R^{2.7}\ B_{_T}^{3.7}}
    {q^{3.1}\ A^{3.53}},
    \ee
where $H$ is the ratio of the actual and predicted confinement times.  Notice, that the result from Eq. (\ref{1.04.46}) matches the result from Eq. (\ref{1.04.37}) very well.  The author of \cite{R3B4v02} does not consider $\kappa$.

\subsection{Conductive power loss for fusion reactors}
Below, we summarize conductive power loss from (\ref{1.04.37}), (\ref{1.04.39}), (\ref{1.04.41}), and (\ref{1.04.44}):
    \be
    \label{1.04.47}
    \begin{split}
    (a) \qquad
    P_{_{\text{Tokamak}}} \propto &
    \Big\{ B_{_T}^{0.32} R^{0.32} \Big\}
    \Big[ \hat{S}_{_F}^{-1.10} \kappa^{0.71} A^{-0.48} \Big]
    \Big(\beta_{_N}^{1.90} q^{1.10}  T^{1.32} \Big)\\
    (b) \qquad
    P_{_{\text{Tokamak}}} \propto &
    \Big\{B_{_T}^{0.43} R^{0.43} \Big\}
    \Big[ \hat{S}_{_F}^{-0.71} \kappa^{0.46} A^{-0.60} \Big]
    \Big( \beta_{_N}^{1.74} q^{0.71}  T^{1.11} \Big) \\
    (c) \qquad
    P_{_{\text{Tokamak}}} \propto &
    \Big[ \hat{S}_{_F}^{-1} \kappa^{0.5}\Big]
    \Big( \beta_{_N}\ q\ T\Big) \\
    (d) \qquad
    P_{_{\text{Stellarator}}} \propto &
    \Big\{B_{_T}^{0.13} R^{0.08}\Big\}
    \Big[ A^{0.85}\Big]
    \Big(\beta^{1.15} q^{1.05} T^{1.41}\Big).
    \end{split}
    \ee
Terms related to reactor size and magnetic field are in curly brackets.  These terms are most important for fusion power.  Terms related to reactor shape are in square brackets.  Terms related to reactor operation are in round brackets.

First, we discuss dependence of conductive power loss on reactor fusion power.
Neutron, Bremsstrahlung, and synchrotron radiation losses are proportional to fusion power.  Dependence of conductive power loss on fusion power is much weaker and much more complicated.

From (\ref{1.04.45}) it follows that fusion reactor power is proportional to $B_{_T}^4\ R^3$.
The model ``IPB98(y,2)" predicts conductive power loss in Tokamaks being proportional to
$B_{_T}^{0.32} R^{0.32}$ as given in (\ref{1.04.47} (a)).  This is approximately proportional to
$P_{_{\text{fusion}}}^{0.09}$.
A more recent model given in (\ref{1.04.38}) predicts conductive power loss in Tokamaks being proportional to $B_{_T}^{0.43} R^{0.43}$ as given in (\ref{1.04.47} (b)).  This is approximately proportional to $P_{_{\text{fusion}}}^{0.13}$.
A theoretical model given in (\ref{1.04.40}) predicts conductive power loss in Tokamaks being independent of $B_{_T}$ and $R$ as given in (\ref{1.04.47} (c)).
The model ``ISS04v3" predicts conductive power loss in Stellarators being proportional to
$B_{_T}^{0.13} R^{0.08}$ as given in (\ref{1.04.47} (d)).  This is approximately proportional to $P_{_{\text{fusion}}}^{0.03}$.
Theoretical results from Covaliu's version of Tang's model predict conductive power loss in Tokamaks and Stellarators being independent of $B_{_T}$ and $R$ \cite[p. 17]{T1}.  Other sources predict steeper dependence of conductive power loss on fusion power.  Costly predicts dependence proportional to $P_{_{\text{fusion}}}^{0.25}$ \cite{R3B4v03}.

Below, we tabulate conduction heat loss at steady-state operation for several built and proposed reactors
\begin{center}
  \begin{tabular}{|l|l|l|l|l|l|l|l|l|l|l|}
  \hline
  Reactor   & Type     & Status   & $P_{_{\text{Fusion}}}$ & $P_{_{\text{Conduction}}}$ & Source \\
            &          &          & $MW$                   & $MW$                       &        \\
  \hline
%  FNS-ST    & D -- T      & Proposed & 0.5    & 14.9 & \cite[p. 10]{STCens01}  \\
  TFTR      & D -- T      & Built    & 9.5    & 20   & \cite[p. 1384]{EConf05} \\
  JET       & D -- T      & Built    & 15.7   & 18.6 & \cite[p. 1384]{EConf05} \\
  IGNITOR   & D -- T      & Proposed & 50     & 20.5 & \cite[p. 5]{EConf06}    \\
  IGNITOR   & D -- T      & Proposed & 75.1   & 39.7 & \cite[p. 1384]{EConf05} \\
  IGNITOR   & D -- T      & Proposed & 95     & 19.4 & \cite[p. 5]{EConf06}    \\
  FIRE      & D -- T      & Proposed & 149    & 33.3 & \cite[p. 1384]{EConf05} \\
  IGNITOR   & D -- T      & Proposed & 155    & 38.5 & \cite[p. 5]{EConf06}    \\
  ITER      & D -- T      & Proposed & 404    & 94.3 & \cite[p. 1384]{EConf05} \\
  Big ITER  & D -- T      & Proposed & 1,500  & 182  & \cite[p. 7]{BigITER}    \\
  \hline
\end{tabular}
  \captionof{table}{Conduction heat loss at steady-state operation for several built and proposed reactors} \label{1.0T10}
\end{center}
Notice that in Table \ref{1.0T10} above, more than one version of IGNITOR reactor is mentioned.  Even in one version, different power levels are likely to appear at different times.  Based on Table \ref{1.0T10} above, conductive power loss is proportional to $P_{_{\text{fusion}}}^{0.43}$.

Overall, dependence of conductive power loss on fusion power is weak -- ranging from $P_{_{\text{fusion}}}^0$ to $P_{_{\text{fusion}}}^{0.43}$.  Accurate data will be obtained only when large Tokamaks and fusion power plants are built.  More powerful reactors lose much lower fraction of fusion power by conduction than the less powerful ones.  This sets a lower power limit for thermonuclear reactors.  For a deuterium -- tritium reactor, the minimal thermal power is about 200 $MW$ \cite{SmallT}.  Minimum power for deuterium -- $^3$He fusion reactor is 20 times higher or 4 $GW$ \cite[p.44]{Spheromak04}.  Tokamaks of lower power should be possible and have been designed, but they are likely to encounter technical problems.

Second, we discuss dependence of conductive power loss on reactor shape.  Eq. (\ref{1.04.47}) parts (a), (b), and (c) present three predicted types of dependence on square bracketed parts:
    \be
    \label{1.04.48}
    \begin{split}
    \hat{S}_{_F}^{-1.10} \kappa^{0.71} A^{-0.48}&=
    S_{_F}^{-1.10} \kappa^{0.71} A^{-1.58}, \\
    \hat{S}_{_F}^{-0.71} \kappa^{0.46} A^{-0.60}&=
    S_{_F}^{-0.71} \kappa^{0.46} A^{-1.31}
    , \\
    \hat{S}_{_F}^{-1} \kappa^{0.5}&=S_{_F}^{-1} \kappa^{0.5} A^{-1}.
    \end{split}
    \ee
The shape factor approximated by (\ref{1.04.17}) drastically grows for lower values of $A$, thus small aspect ratio Tokamaks have lower conductive power loss.  Optimal elongation is $2 \le \kappa \le 3.5$ \cite[p.14]{ST19}.

Second, we discuss dependence of conductive power loss on reactor operating conditions.  Eq. (\ref{1.04.47})  presents four predicted types of dependence on round bracketed parts:
    \be
    \label{1.04.49}
    \beta_{_N}^{1.90} q^{1.10}  T^{1.32}, \qquad
    \beta_{_N}^{1.74} q^{0.71}  T^{1.11}, \qquad
    \beta_{_N}\ q\ T, \qquad
    \beta^{1.15} q^{1.05} T^{1.41}.
    \ee
Conductive power loss increases with normalized $\beta$, safety factor, and operating temperature.

Spherical Tokamaks which consume more power than they produce can be very small.  FNS-ST Tokamak has major radius $R=0.5\ m$, minor radius $a=0.3\ m$, and toroidal field $B_{_T}=1.5\ Tesla$.   It has heating power of 15 $MW$ and fusion power of 500 $kW$.
ST-CTF reactor has major radius $R=0.81\ m$, minor radius $a=0.52\ m$, and toroidal field $B_{_T}=2.6\ Tesla$.  It has heating power of 44 $MW$ and fusion power of 35 $kW$ \cite[p.10]{STCensus01}.  Aforementioned reactors would be useful as neutron sources or as parts of hybrid fission-fusion reactors.

\subsection{Lawson pressure criterion}

Below, we summarize Lawson pressure criterion from (\ref{1.04.37}), (\ref{1.04.39}), (\ref{1.04.41}), and (\ref{1.04.44}):
    \be
    \label{1.04.50}
    \begin{split}
    (a) \qquad C_{_{LPT}} \propto &
    \Big\{ B_{_T}^{3.68} R^{2.68} \Big\}
    \Big[ \hat{S}_{_F}^{3.10}  A^{-3.52} \kappa^{0.29} \Big]
    \Big( \beta_{_N}^{0.10} q^{-3.1}  T^{-1.32} \Big) \\
    (b) \qquad C_{_{LPT}} \propto &
    \Big\{ B_{_T}^{3.57} R^{2.57} \Big\}
    \Big[ \hat{S}_{_F}^{2.71}  A^{-3.4} \kappa^{0.54} \Big]
    \Big( \beta_{_N}^{0.26} q^{-2.71}  T^{-1.11} \Big)\\
    (c) \qquad C_{_{LPT}} \propto &
    \Big\{  B_{_T}^{  4} R^{  3} \Big\}
    \Big[ \hat{S}_{_F}^{3} A^{ -4} \kappa^{0.5} \Big]
    \Big( \beta_{_N}\ q^{ -3}  T^{ -1} \Big)\\
    (d) \qquad C_{_{LPS}} \propto &
    \Big\{ B_{_T}^{3.87} R^{2.92} \Big\}
    \Big[ A^{-2.85} \Big]
    \Big( \beta^{0.85} q^{-1.05} T^{-1.41} \Big).
    \end{split}
    \ee
Rewriting (\ref{1.04.50}) in terms of $S_{_F}$, we have
    \be
    \label{1.04.51}
    \begin{split}
    (a) \qquad C_{_{LPT}} \propto &
    \Big\{ B_{_T}^{3.68} R^{2.68} \Big\}
    \Big[ S_{_F}^{3.10}  A^{-0.42} \kappa^{0.29} \Big]
    \Big( \beta_{_N}^{0.10} q^{-3.1}  T^{-1.32} \Big) \\
    (b) \qquad C_{_{LPT}} \propto &
    \Big\{ B_{_T}^{3.57} R^{2.57} \Big\}
    \Big[ S_{_F}^{2.71}  A^{-0.69} \kappa^{0.54} \Big]
    \Big( \beta_{_N}^{0.26} q^{-2.71}  T^{-1.11} \Big)\\
    (c) \qquad C_{_{LPT}} \propto &
    \Big\{  B_{_T}^{  4} R^{  3} \Big\}
    \Big[ S_{_F}^{3} A^{ -1} \kappa^{0.5} \Big]
    \Big( \beta_{_N}\ q^{ -3}  T^{ -1} \Big)\\
    (d) \qquad C_{_{LPS}} \propto &
    \Big\{ B_{_T}^{3.87} R^{2.92} \Big\}
    \Big[ A^{-2.85} \Big]
    \Big( \beta^{0.85} q^{-1.05} T^{-1.41} \Big).
    \end{split}
    \ee
The notations are similar to those used in the last subsection.  Terms related to reactor size and magnetic field are in curly brackets.  These terms are most important for fusion power.  Terms related to reactor shape are in square brackets.  Terms related to reactor operation are in round brackets.

Lawson pressure criterion is almost linearly proportional to reactor fusion power.  From (\ref{1.04.45}) it follows that fusion reactor power is proportional to $B_{_T}^4\ R^3$.  Expressions in curly brackets in Eq. (\ref{1.04.51}) are almost similar.

According to Eq. (\ref{1.04.51}) parts (a), (b), and (c), Lawson pressure criterion of Tokamaks is almost proportional to $S_{_F}^{3}$.  As we see from approximation (\ref{1.04.17}) for the shape factor, reactors with low aspect ratio and high elongation have the highest shape factors.

As we see from expressions in round brackets in (\ref{1.04.51}), running a reactor at low safety factor and lower temperature increases the Lawson pressure criterion.  Nevertheless, both approaches create technical problems.

Define the \textbf{reactor criterion} as
    \be
    \label{1.04.52}
    \mathcal{R}_{_C}=S_{_F}^{3} B_{_T}^{4} R^{3}.
    \ee
Reactor criterion has units of $Tesla^4\ m^3$.  As we see from (\ref{1.04.51}), the Lawson pressure criterion for Tokamaks and Stellarators is proportional to the reactor criterion to the power of 0.9 to 0.97.  The Tokamak fusion power is proportional to \cite[p.1]{SFact01}:
    \be
    \label{1.04.53}
    P_{_{_{\text{fusion}}}} \propto
    S_{_F}^2 B_{_T}^{4} R^{3} \beta_{_N}^2 q^{-2}=
    \frac{\beta_{_N}^2}{S_{_F} q^2}\
    \mathcal{R}_{_C}.
    \ee

\subsection{The Reactor criterion for reactors similar to ITER}

Below we estimate $\mathcal{R}_{_C}$ required for fusion reactors similar to International Thermonuclear Experimental Reactor (ITER) and using different types of fuel.  First, we calculate $\mathcal{R}_{_C}$ required for deuterium -- tritium fusion reactor.
ITER has toroidal magnetic field $B_{_T}=5.3\ Tesla$, major radius $R=6.2\ m$, and shape factor $S_{_F}=4.2$ \cite[p.243]{FusBk2}.
Hence, it has the Reactor Criterion
    \be
    \label{1.04.54}
    \mathcal{R}_{_C}\big(\text{ITER} \big)=1.4 \cdot 10^7\ Tesla^4\ m^3.
    \ee
Notice, that ITER reactor does not achieve ignition.
A working nuclear power plant should have an even higher $\mathcal{R}_{_C}$, even though it is not a requirement.  Ignition should be achieved for $R=7.5$ \cite[p.14]{Lawson1}.
Big ITER has the major radius $R=8.14\ m$,  toroidal magnetic field
$B_{_T}=5.68\ Tesla$, and shape factor $S_{_F}=3.9$.
This reactor has fusion power of 1.5 $GW$ \cite[p.7]{BigITER}.  Big ITER has
    \be
    \label{1.04.55}
    \mathcal{R}_{_C}\big(\text{Big ITER} \big)=3.3 \cdot 10^7\ Tesla^4\ m^3.
    \ee
%A proposed fusion power plant called K-DEMO has $R=6.8\ m$, $a=2.1\ m$, $B_{_T}=7.4\ Tesla$, and hence $\mathcal{R}_{_C}=8.58 \cdot 10^3\ Tesla^4\ m^3$.  This power plant would generate 3 $GW$ thermal fusion power \cite[p.2]{SCMg1}.

Below we calculate $\mathcal{R}_{_C}$ needed for a reactor similar to ITER which uses deuterium -- $^3$He fusion reaction. Assuming that the values of $F_{_{\text{Reactor}}}$ and  $F_{_{\text{RO}}}$ are equal for similar reactors regardless of the fuel, we derive the following from (\ref{1.04.37}), (\ref{1.04.39}), (\ref{1.04.41}), and (\ref{1.04.44}):
    \be
    \label{1.04.56}
    \begin{split}
    \mathcal{R}_{_C}& \propto
    C_{_{LPT}}^{1.1}\ T^{1.44}\ \overline{M}^{-0.67}\ (\overline{Z}+1)^{1.45}
    \qquad \text{for Tokamaks},
    \\
    \mathcal{R}_{_C}& \propto C_{_{LPS}}^{1.1}\ T^{1.45}\ (\overline{Z}+1)^{1.45} \hskip2.2cm \text{for Stellarators}.
    \end{split}
    \ee
If $\mathcal{R}_{_{C1}}$ is a reactor criterion for a Tokamak using fuel 1, then a similar Tokamak using fuel 2 would need the following reactor criterion derived from (\ref{1.04.56}):
    \be
    \label{1.04.57}
    \begin{split}
    \mathcal{R}_{_{C2}}&=\mathcal{R}_{_{C1}}\
    \frac{
    \left[C_{_{LPT}}^{1.1}\ T^{1.44}\ \overline{M}^{-0.67}\ (\overline{Z}+1)^{1.45}
    \right]_{\text{Fuel 2}}}
    {\left[C_{_{LPT}}^{1.1}\ T^{1.44}\ \overline{M}^{-0.67}\ (\overline{Z}+1)^{1.45}
    \right]_{\text{Fuel 1}}}.
    \end{split}
    \ee

Deuterium -- $^3$He reactor would need Lawson Criterion 27 times higher than a deuterium -- tritium reactor \cite[p.81]{AF01}.  As we see from Table \ref{1.0T03}, a deuterium -- $^3$He reactor would have operating plasma temperature 4.3 times higher than a deuterium -- tritium reactor.
Plasma consisting of deuterium and $^3$He would have an average nuclear charge of $Z=1.5$.  Plasma consisting of deuterium and tritium would have an average nuclear charge of $Z=1$.
Both plasmas will have the same average nuclear mass.

Substituting the data from the last paragraph into (\ref{1.04.57}) we conclude that the reactor criterion for a deuterium -- $^3$He reactor should be about 424 times higher than the reactor criterion for a similar deuterium -- tritium reactor.
The reactor criterion for a deuterium -- $^3$He reactor with proportions similar to Big ITER is:
    \be
    \label{1.04.58}
    \mathcal{R}_{_C}\Big(\ ^2\text{H}-^3\text{He} \Big)=1.4 \cdot 10^{10}\ Tesla^4\ m^3.
    \ee
A deuterium -- $^3$He Tokamak has a very high reactor criterion.  Thus, aneutronic fusion has almost no prospect within the next 50 years or perhaps for the rest of the Century.  Nevertheless, aneutronic fusion reactors can play a major role in Solar System Colonization and Deep Space propulsion in the next Century.

Apollo is a design of a deuterium -- $^3$He Tokamak.  This Tokamak has a toroidal magnetic field of $B_{_T}=10.9\ Tesla$, major radius $R=7.9\ m$ and shape factor $S_{_F}=5.1$.  Apollo Tokamak has fusion power of 2.1 $GW$ \cite{AN03}.  The reactor coefficient of Apollo Tokamak is
    \be
    \label{1.04.59}
    \mathcal{R}_{_C}\Big(\ ^2\text{H}-^3\text{He},\ \text{Apollo} \Big)=9.2 \cdot 10^8\ Tesla^4\ m^3.
    \ee
This coefficient is 15 times lower than the coefficient predicted by (\ref{1.04.58}).  This model may be too optimistic.  Moreover, as we argue in Subsection {\rr 5.3}, a non-spherical Tokamak for deuterium -- $^3$He fusion may be impossible.

\subsection{Reactor criteria for spherical deuterium -- $^3$He Tokamaks}

A Spherical deuterium -- $^3$He Tokamak has been designed in a great detail.  This Tokamak has toroidal magnetic field of $B_{_T}=2.7\ Tesla$, major radius
$R=8\ m$, unreasonably high shape factor $S_{_F}=124$, and plasma volume of 17,900 $m^3$ \cite{AF01,ANST01}.
The reactor coefficient of this Tokamak is
    \be
    \label{1.04.60}
    \mathcal{R}_{_C}\Big(\ ^2\text{H}-^3\text{He},\ \text{spherical} \Big)=5.1 \cdot 10^{10}\ Tesla^4\ m^3.
    \ee
This coefficient is 3.6 times higher than the coefficient predicted by (\ref{1.04.58}).

Another proposed deuterium -- $^3$He Tokamak is the GA Project 4437 Tokamak.  This Tokamak has toroidal magnetic field of $B_{_T}=2.7\ Tesla$ and major radius
$R=9.45\ m$.  The Tokamak has aspect ratio $A=1.4$, plasma elongation $\kappa=2.5$, and triangularity $\delta=0.8$ \cite[p.45]{Spheromak04}.  Using (\ref{1.04.17}), we estimate the shape factor of $S_{_F}=60$.
The reactor coefficient of GA Project 4437 Tokamak is
    \be
    \label{1.04.61}
    \mathcal{R}_{_C}\Big(\ ^2\text{H}-^3\text{He},\ \text{spherical} \Big)=9.7 \cdot 10^9\ Tesla^4\ m^3.
    \ee
In Section {\rr 6.2}, we also mention a Tokamak of our design -- Spheromak2100.  This Tokamak has a toroidal magnetic field of $B_{_T}=4.1\ Tesla$ and major radius $R=9.45\ m$.  This Tokamak would have the same shape and dimensions as GA Project 4437 Tokamak, thus $S_{_F}=60$.
The reactor coefficient of this Tokamak is
    \be
    \label{1.04.62}
    \mathcal{R}_{_C}\Big(\ ^2\text{H}-^3\text{He},\ \text{spherical} \Big)=5.2 \cdot 10^{10}\ Tesla^4\ m^3.
    \ee

\section{Ways of improving Lawson pressure criterion}
Improving Lawson pressure criterion for a Tokamak or Stellarator reactors is the main objective in achieving commercial deuterium -- tritium fusion power generation.  It is vital for building a deuterium -- $^3$He Tokamaks or Stellarators, as these reactors require 27 times higher Lawson Criterion.

Recall the expression pressure criteria (\ref{1.04.51}) for Tokamaks and Stellerators:
    \be
    \label{1.05.01}
    \begin{split}
    (a) \qquad C_{_{LPT}} \propto &
    \Big\{ B_{_T}^{3.68} R^{2.68} \Big\}
    \Big[ S_{_F}^{3.10}  A^{-0.42} \kappa^{0.29} \Big]
    \Big( \beta_{_N}^{0.10} q^{-3.1}  T^{-1.32} \Big) \\
    (b) \qquad C_{_{LPT}} \propto &
    \Big\{ B_{_T}^{3.57} R^{2.57} \Big\}
    \Big[ S_{_F}^{2.71}  A^{-0.69} \kappa^{0.54} \Big]
    \Big( \beta_{_N}^{0.26} q^{-2.71}  T^{-1.11} \Big)\\
    (c) \qquad C_{_{LPT}} \propto &
    \Big\{  B_{_T}^{  4} R^{  3} \Big\}
    \Big[ S_{_F}^{3} A^{ -1} \kappa^{0.5} \Big]
    \Big( \beta_{_N}\ q^{ -3}  T^{ -1} \Big)\\
    (d) \qquad C_{_{LPS}} \propto &
    \Big\{ B_{_T}^{3.87} R^{2.92} \Big\}
    \Big[ A^{-2.85} \Big]
    \Big( \beta^{0.85} q^{-1.05} T^{-1.41} \Big).
    \end{split}
    \ee
There are four approaches to improving Lawson pressure criterion of a fusion reactor.
First, we can increase reactor torus major radius $R$.
Second, we can increase reactor shape factor of $S_{_F}$.
Third, we can increase toroidal magnetic field $B_{_T}$.
Fourth, we can run the reactor at higher $\beta$ and lower safety factor $q$.

\subsection{Greenwald density limit}

Lawson pressure criterion of Tokamaks and Stellarators can be improved by increasing $R$, $B_{_T}$, and $\beta$.  These improvements run up against a limit which we discuss in this Subsection.  Based on decades of observation, it has been determined that the density of nuclei in plasma can not exceed the \textbf{Greenwald density limit} given by \cite[p.52]{FusBk1}
    \be
    \label{1.05.02}
    \frac{n_{_G}}{10^{20}\ m^{-3}}=
    \frac{1}{\pi}\ \left(\frac{I}{10^6\ A}\right) \left(\frac{a}{1\ m} \right)^{-2}=
    \frac{A^2}{\pi}\ \left(\frac{I}{10^6\ A}\right) \left(\frac{R}{1\ m} \right)^{-2},
    \ee
where $MA$ denotes Mega Amperes.
Substituting (\ref{1.02.03}) into (\ref{1.05.02}) we obtain \textbf{Greenwald pressure limit}
    \be
    \label{1.05.03}
    P_{_G}=0.0510\ bar\ A^2\ \big(1+\overline{Z}\big)\ \left(\frac{T}{1\ keV}\right)
    \ \left(\frac{I}{10^6\ A}\right) \left(\frac{R}{1\ m} \right)^{-2}.
    \ee

Using (\ref{1.02.04}) and the definition of $\beta$, we obtain plasma pressure:
    \be
    \label{1.05.04}
    P_{_g}=3.98\ bar\ \left(\frac{B_{_T}}{1\ Tesla}\right)^2 \beta
    \ee
The \textbf{Greenwald ratio} is the ratio of plasma pressure to Greenwald pressure limit.  It is given by
    \be
    \label{1.05.05}
    R_{_G}=\frac{P_{_g}}{P_{_G}}=
    \frac{78.0\  \beta}{\big(1+\overline{Z}\big)\ A^2}
    \left(\frac{B_{_T}}{1\ Tesla}\right)^2\
    \left(\frac{R}{1\ m}\right)^2\
    \ \left(\frac{I}{10^6\ A}\right)^{-1}\
    \left(\frac{T}{1\ keV}\right)^{-1}.
    \ee
Normalized $\beta$ is defined as \cite[p.9]{FusBk3}:
    \be
    \label{1.05.06}
    \beta_{_N}=\frac{100\ \beta}{A}\ \left(\frac{I}{10^6\ A}\right)^{-1}
    \left(\frac{R}{1\ m}\right) \left(\frac{B_{_T}}{1\ Tesla}\right).
    \ee
From Eq. (\ref{1.05.06}), we obtain
    \be
    \label{1.05.07}
    \beta=\frac{A\ \beta_{_N}}{100}\ \left(\frac{I}{10^6\ A}\right)
    \left(\frac{R}{1\ m}\right)^{-1} \left(\frac{B_{_T}}{1\ Tesla}\right)^{-1}.
    \ee
Substituting (\ref{1.05.07}) into (\ref{1.05.05}), we obtain
    \be
    \label{1.05.08}
    R_{_G}=
    \frac{0.78\  \beta_{_N}}{\big(1+\overline{Z}\big)\ A}
    \left(\frac{B_{_T}}{1\ Tesla}\right)\
    \left(\frac{R}{1\ m}\right)\
    \left(\frac{T}{1\ keV}\right)^{-1}.
    \ee

In a working reactor, the Greenwald ratio should be below 0.7 and must be below 0.8 \cite[p.57]{FusBk1}.  Greenwald ratios for several proposed reactors are presented in Table \ref{1.0T11} below.
\begin{center}
  \begin{tabular}{|l|l|l|l|l|l|l|l|l|l|l|l|}
  \hline
  Reactor         & Reaction & Thermal     & $R_{_G}$ & Source  \\
                  &          & Power, $GW$ &          &         \\
  \hline
  ITER            & D-T      & 0.4         & 0.84     & \cite[p.243]{FusBk2} \\
  Big ITER        & D-T      & 1.5         & 0.75     & \cite[p.7]{BigITER}  \\
  STPP            & D-T      & 3.1         & 0.63     & \cite[p.16]{FusBk2}  \\
  \hline
                  & D-$^3$He &             & 0.39     & \cite{AF01}          \\
  GA Project 4437 & D-$^3$He & 11.0        & 0.48     & \cite[p.45]{Spheromak04} \\
  Apollo          & D-$^3$He & 2.14        & 0.55     & \cite[p.2]{AN03}     \\
  \hline
\end{tabular}
  \captionof{table}{Greenwald ratios for proposed reactors} \label{1.0T11}
\end{center}

As we see from (\ref{1.05.08}) above, the Greenwald limit presents an effective limit to increasing the reactor's major radius and toroidal magnetic field.  Recalling (\ref{1.04.45}), the reactor's fusion power is proportional to $R^3 B_{_T}^4$.  Thus, the Greenwald limit presents an effective limit to reactor power.  Power of deuterium -- tritium reactors is likely to have an upper limit of about 5 $GW$ to 10 $GW$.  Power of deuterium -- $^3$He reactors is likely to have an upper limit of about 50 $GW$ to 100 $GW$.  Accurate determination of these limits remains an open problem.

From (\ref{1.05.08}) above, it follows that a reactor operating at thermal power close to the upper limit should operate at high temperature.  For a deuterium-tritium reactor, a good operating temperature should be about 25 $keV$.  As we see from Table \ref{1.0T09}, at higher temperatures, energy losses due to synchrotron radiation from plasma become too high.

\subsection{Increasing reactor torus major radius $R$}

Historically, the major radius (R) of the Tokamak Stellarator toruses grew over time.
In 1958, Soviet T1 Toakamak had $R=0.67\ m$.
In 1962, Soviet T3 Tokamak had $R=1.0\ m$.
In 1975, Soviet T10 Tokamak had $R=1.5\ m$ \cite{TokUSSR}.
Joint European Torus Tokamak built in 1984 had $R=2.96\ m$.  International Thermonuclear Experimental Reactor (ITER) will have $R=6.2\ m$ \cite[p. 19]{TProgress01}.
Original (1998) design for ITER had $R=8.14\ m$ \cite[p. 14]{Hist01}.  This design was abandoned due to high cost.

\subsection{Increase reactor shape factor of $S_{_F}$.}

This is accomplished by educing reactor torus aspect ratio $A$.
According to Eq. (\ref{1.05.01}), the Lawson pressure criterion of Tokamak is proportional to $S_{_F}^3$.  As we see from (\ref{1.04.17}), reducing aspect ratio greatly increases the shape factor.  Also, plasma elongation should be high.  Generally, spherical Tokamaks have $2.4 \le \kappa \le 3.5$.  Spherical Tokamaks should have plasma triangularity of at least $\delta=0.5$ \cite[p.14]{ST19}.
For Stellarators, aspect ratio does not play as important role as it does for Tokamaks.

A \textbf{Spherical Tokamak} is "a tokamak with the central region reduced to the
minimum size possible" \cite[p. 59]{FusBk1}.  The torus looks like a sphere with a pencil-shaped region cut from the middle.  A center-post carrying very high electric current makes up the pencil-shaped region.  "The National Spherical Torus Experiment has achieved $\beta=0.3-0.4$  and has consistently achieved energy confinement times 2–3 times larger than predicted by conventional Tokamak correlations" \cite[p. 59]{FusBk1}.  Theoretically, $\beta=0.5$ should be sustainable in spherical Tokamaks \cite[p. 376]{Freidberg}.  Spherical Tokamaks have aspect ratios between 1.4 to 1.8 \cite[p. 8]{STK01}.  This is about half of ITER's ratio of 3.1.

One advantages of using spherical Tokamaks is that some designs do not use superconductive toroidal field coil magnets \cite[p. 17]{ST19}.  If such Tokamaks are possible, their cost would be significantly reduced.

For spherical Tokamaks, shape factors as high as $S_{_F}=41$ have been achieved \cite[p. 268]{Tokamak}.  National Spherical Torus Experiment (NSTX) is a Tokamak with major radius $R=0.85\ m$, minor radius $0.6\ m$, aspect ratio $A=1.42$, and elongation $\kappa=2.2$.  In different runs, NSTX has values of $30 \le S_{_F} \le 38$.  This reactor has sustained $\beta$ of 0.22 to 0.25.  NSTX discharges have $\beta_{_N}$ between 5 and 6 \cite{SFact02}.

A spherical Tokamak reactor is illustrated in Figure \ref{1.0F02} below.  This Tokamak has $A=1.6$, $\kappa=2.3$, and $\delta \approx 0.9$.  According to Eq. (\ref{1.04.17}), this Tokamak should have $S_{_F}=33$.
\begin{center}
\includegraphics[width=12cm,height=12cm]{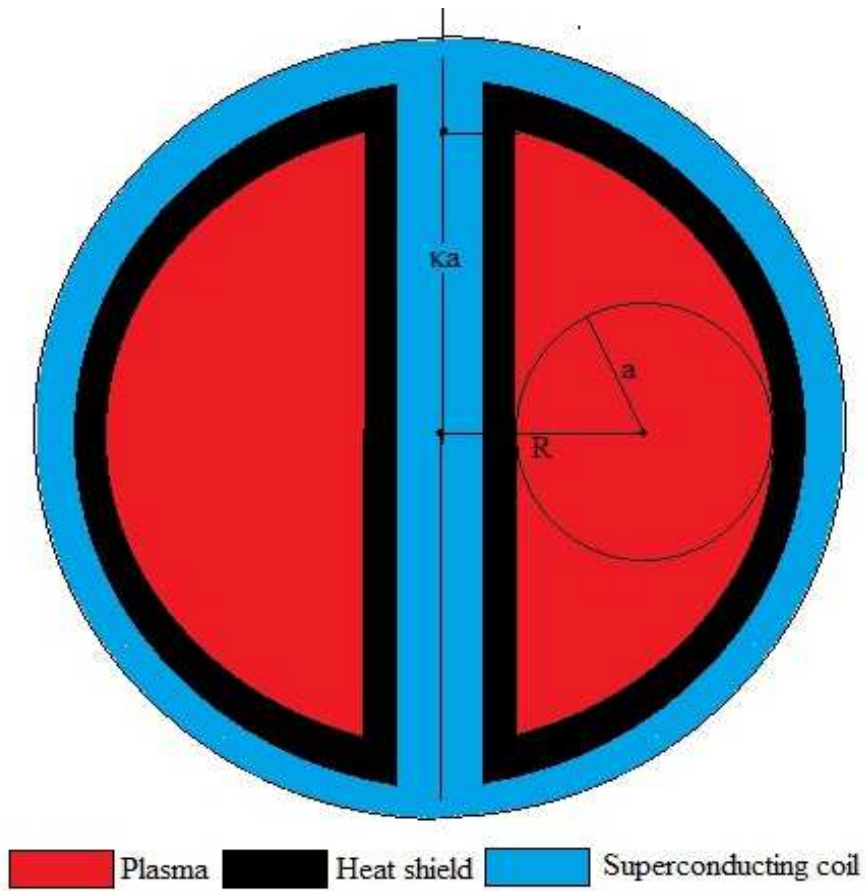}
\captionof{figure}{Spherical Tokamak cross-section \label{1.0F02}}
\end{center}

Spherical Tokamaks do present very significant engineering difficulties.  The center-post is exposed to very high loads of heat and neutron radiation.  ARIES Spherical Tokamak concept envisions heat flux of 6.4 $MW/m^2$ to the middle section of the center-post.  About 80\% of this heat is carried by neutron radiation \cite[p.31]{STK01}.

Another problem for spherical Tokamaks is very high magnetic field at the superconducting wires running within the center-post.  As we see from Table \ref{1.0T11} below, the ratio of magnetic field at the wires on the inner circle of the torus to $B_{_T}$ increases with decreasing $A$.
\begin{center}
  \begin{tabular}{|c|r|r|l|l|l|l|l|l|l|l|l|}
  \hline
  Reactor & Status    & $B_{_T}$ & $B_{_{\text{Max}}}$ &Field & A       & Source              \\
          &           & $Tesla$  & $Tesla$  & Ratio           &         &                     \\
  \hline
  Aries-ST& Project   & 2.14     & 7.6      &  3.55         & 1.6     & \cite[p.246]{FusBk2} \\
  Vector  & Project   & 4.7      & 19.6     &  4.17         & 2.0     & \cite[p.246]{FusBk2} \\
  SlimCS  & Project   & 6.0      & 16.4     &  2.73         & 2.6     & \cite[p.246]{FusBk2} \\
  JT-60SA & Completed & 2.68     & 6.5      &  2.43         & 2.7     & \cite[p.21]{SCMS01}  \\
  ITER    & UC        & 5.3      & 13.5     &  2.55         & 3.1     & \cite[p.117]{FusBk1} \\
  TS      & Completed & 4.5      & 9.3      &  2.07         & 3.2     & \cite[p.18]{SCMS01}  \\
  SABR    & Project   & 5.7      & 13.5     &  2.37         & 3.4     & \cite[p.231]{FusBk1} \\
  KSTAR   & Completed & 3.5      & 7.2      &  2.06         & 3.6     & \cite[p.20]{SCMS01}  \\
  Aries-RS& Project   &   8      & 16       &  2.00         & 4.0     & \cite[p.227]{FusBk1} \\
  EAST    & Completed & 3.5      & 5.8      &  1.66         & 4.3     & \cite[p.19]{SCMS01}  \\
  Aries-I & Project   &  11      & 19       &  1.73         & 4.5     & \cite[p.227]{FusBk1} \\
  SST-1   & Completed & 3.0      & 5.1      &  1.70         & 5.5     & \cite[p.20]{SCMS01}  \\
  \hline
\end{tabular}
  \captionof{table}{Tokamak magnetic field} \label{1.0T12}
\end{center}

It is possible that spherical Tokamak reactors would present new and unexpected problems.  This happens to every technology on its way to maturity.  For instance, Thompson's and Blackman's 1946 patent envisioned a deuterium-deuterium fusion reactor with major radius $R=1.3\ m$, minor radius $a=0.3\ m$ and operating temperature of 500 $keV$ \cite[p.3]{Spheromak02}.  Energy confinement time of 65 $s$ was an overestimate by a factor of at least $10^5$.  Given that they had no experimental data, their mistake is understandable.  It is possible that our understanding of fully working Tokamaks which have not been built yet is equally flawed.

Recalling (\ref{1.03.15}), for a deuterium -- $^3$He reactor, the synchrotron power fraction is proportional to $\beta^{-1.5}$.  As we see from Table \ref{1.0T13}, synchrotron power fraction is substantial even for $\beta=0.3$.  In order to operate, a deuterium -- $^3$He reactor must have $\beta \ge 0.3$.  Such high values of $\beta$ are possible only for spherical Tokamaks.  Thus, spherical Tokamaks are required for deuterium -- $^3$He fusion.

\subsection{Increasing toroidal magnetic field $B_{_T}$}

According to Eq. (\ref{1.05.01}), the Lawson pressure criterion of Tokamak is proportional to about $B^{3.7}$, and the Lawson pressure criterion of Stellarator is proportional to $B^{3.87}$.  Moreover, fusion reactor power is proportional to $B^4$.  Thus, increasing the magnetic field is very important.

In all working fusion reactors, magnetic fields would be produced by superconducting electromagnets.  In order for an electric conductor to be superconducting, it must be kept below critical temperature $T_{_{\text{critical}}}$.  It also has to be kept below temperature-dependent critical magnetic field $B_{_{\text{critical}}}(T)$.  Generally, $B_{_{\text{critical}}}(T)$ is a strongly decreasing function of temperature, thus a superconductor which has to sustain a strong magnetic field must be kept well below critical temperature.  ITER uses Nb$_3$Sn superconducting wires at 4.5 $^o$K temperature.  The gross weight of all coils is 9,677 $tons$ \cite[p.23]{SCMS01}.

For a reactor similar to ITER, the magnetic field at the wires on the inner circle of the torus is $2.55\ B_{_T}$.  For a reactor with aspect ratio $A=2$, the magnetic field at the wires on the inner circle of the torus is $3.5\ B_{_T}$ to $3.5\ B_{_T}$ \cite[p.246]{FusBk2}.  Thus, Tokamaks and Stellarators using high magnetic fields must sustain very high magnetic fields on the superconducting wires.

There are two types of superconductors.  The first type is called low temperature superconductor (LTS).  These superconductors are almost always metallic.  Critical magnetic fields for all LTS are at most 23.5 $Tesla$.  These superconductors are already used in magnetic resonance imaging (MRI) and large hadron collider (LHC) magnets \cite[p.1]{HTSTape04}.

The second type of superconductors are high temperature superconductors.  These superconductors have much higher critical temperatures and critical magnetic fields.  These superconductors are ceramic.  They can be somewhat flexible as very thin foil attached to a tape.  REBCO tape is 4 $mm$ wide and .12 $mm$ thick.  The superconducting layer is only 2 $\mu m$ thick \cite[p.2]{HTSTape04}.

High temperature superconductor tape is indispensable for all applications with fields over 25 $Tesla$ and for all applications with operating temperature of 10 $^o$K or higher \cite[p.5]{HTSTape03}.  "High temperature superconductor tape (HTST) can carry high current densities even at field strengths of 30 $Tesla$" \cite[p.1]{HTSTape04}.

High temperature superconductor tape would definitely enable Tokamaks to have toroidal field of $B_{_T}=12\ Tesla$. Superconducting tapes can carry high currents in a 40 $Tesla$ field \cite[p.6]{HField1}.  Thus,  toroidal fields of $B_{_T}=16\ Tesla$ may be possible in a few decades.  A blueprint for a 100 $Tesla$ superconducting magnet has been produced \cite{100T}.

A new class of superconducting materials have been recently discovered.  These materials become superconducting at room temperature at pressures of several million atmospheres.  Some carbon-doped sulfur hydrides exhibit superconductivity at 15 $^o$C at a pressure of 2.7 million atmospheres \cite{RTSC}.  Carbon nanotubes have tensile strength of 1.5 million atmospheres, thus super pressurized superconducting wires may be possible \cite{CNT1}.  Possibly, these wires will have very high critical magnetic field.

\subsection{Decreasing safety factor $q$ and increasing $\beta_{_N}$}
According to Eq. (\ref{1.05.01}), the Lawson pressure criterion of Tokamak is proportional to about $q^{-3}$, and Lawson pressure criterion of Stellarator is proportional to $q^{-1.05}\cdot \beta^{0.85}$.  According to Eq. (\ref{1.04.53}), fusion reactor power is proportional to $\beta_{_N}^2\ q^{-2}$.  According to Eq. (\ref{1.03.15}), the fraction of fusion power lost to synchrotron radiation is proportional to $\beta^{-1.5}$. Since $\beta \propto \beta_{_N}$, synchrotron radiation loss is proportional to $\beta_{_N}^{-1.5}$.
Thus, decreasing safety factor $q$ and increasing $\beta_{_N}$ are important goals.

ITER has $\beta_{_N}=1.77$ \cite[p.243]{FusBk2}. If $\beta_{_N}$ exceeds the \textbf{Troyon limit}, then plasma becomes unstable \cite[p.933]{Tokamak}.  According to Troyon himself, the limit is $\beta_{_N} \le 2.8$ \cite{Troyon}.  According to theoretical calculations, $\beta_{_N} \le 3.5$ \cite[p.332]{Tokamak}.  START spherical Tokamak experiment achieved stability for $\beta_{_N}=6$ and $\beta=0.4$.

Most proposed Tokamaks have safety factors between 3 and 4.  ITER has safety factor $q=3$ \cite[p.243]{FusBk2}.  Some researchers believe that Tokamaks and Stellarators can operate with safety factors $q \le 2$ \cite{SF03}.

Operating at low safety factor and high $\beta_{_N}$ would greatly enhance reactor performance.  Nevertheless, operating in unsafe regimes is likely to cause many reactor shutdowns and accidents.  A fusion reactor accident would not consist of a thermonuclear explosion, but it may break a reactor worth billions.  Moreover, such an accident may release large amount of radioactive material.

\section{Prospects for reactor development}
\subsection{Likely timeline}
In our opinion, deuterium -- tritium fusion power will play a considerable role in Global energy production during the second half of this century.  Deuterium -- $^3$He fusion power is unlikely to play any role on Earth, but it is likely to play a major role during Solar System colonization during the next century.

We are not optimistic about the potential for thermonuclear power in the short time frame.  Over the previous four decades, fusion power research has received very low budget.  Between 1960 and 1974, annual US budget for magnetic confinement fusion was about \$200 million \cite[p.79]{Hist02}.
Between 1975 and 1982, magnetic confinement fusion research in the USA received annual funding of  about \$1.0 billion.  In later years, annual funding for magnetic confinement fusion was continuously decreasing until about 1997 \cite{NFBudget1}.  Between 2000 and 2012, the average annual magnetic confinement fusion funding in the USA was \$300 million to \$400 million \cite{NFBudget2}.
Overall fusion funding based on \cite[p. 7]{LCostF1} is presented in Figure \ref{1.0F03} below:
\begin{center}
\includegraphics[width=16cm]{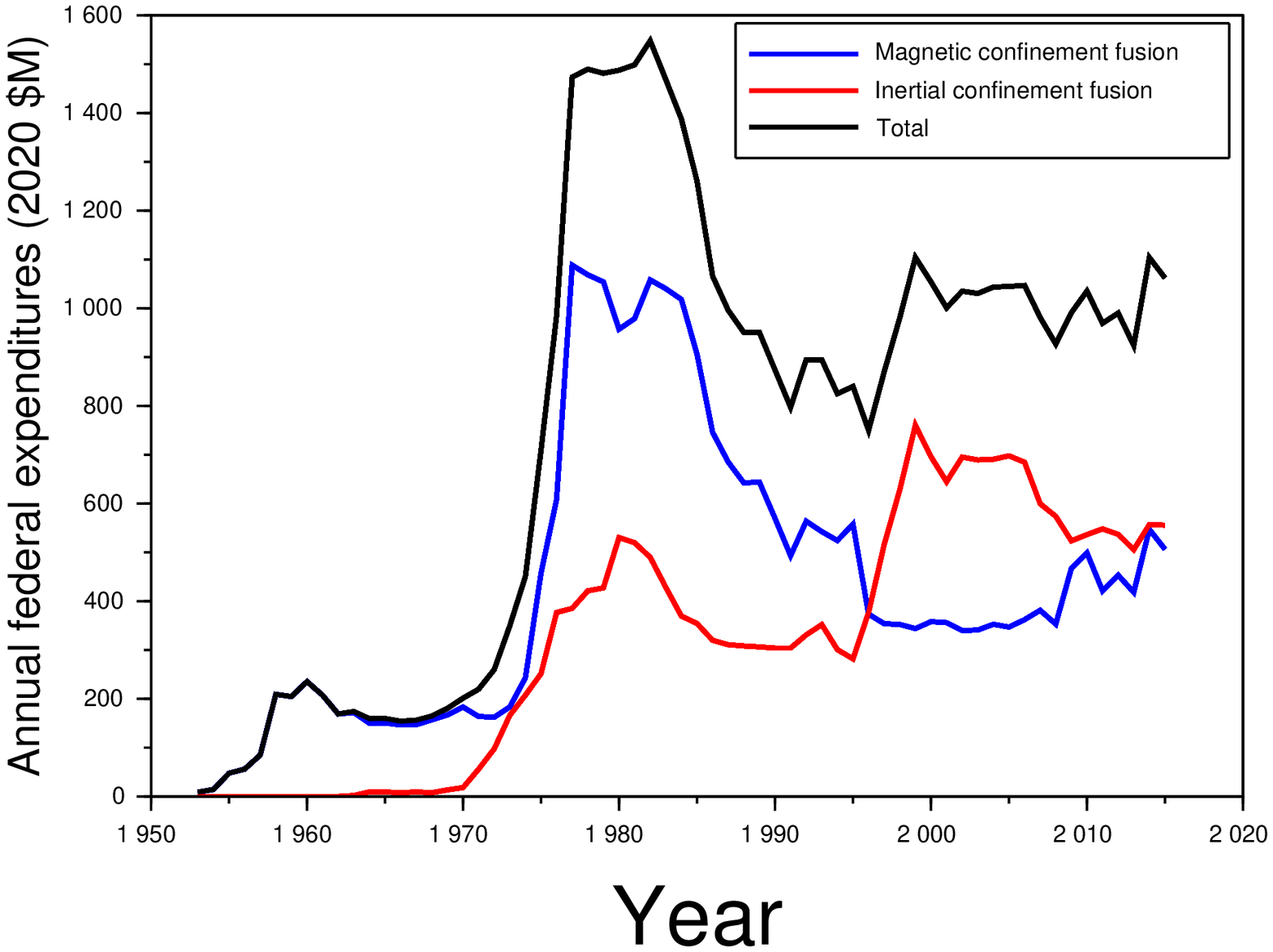}
\captionof{figure}{Federal Fusion Funding \label{1.0F03}}
\end{center}

A 1976 plan for development of magnetic confinement fusion drew several scenarios for funding of magnetic confinement fusion.
Maximum effective effort would have required about \$9 billion per year for 11 years.
The medium effort plan would have required \$4.5 billion per year for 14 years.
The low effort plan would have required \$3 billion per year for 26 years.
Funding of \$1.2 billion per year was predicted to bring no result \cite[p.12]{FPlan}.
In reality, average annual funding for magnetic confinement fusion between 1978 and 2015 has been \$460 million per year.
All amounts mentioned in this subsection are in year 2020 dollars.

Lawson pressure criterion of the newest reactors has experienced phenomenal growth between 1955 and 1997.  Top Lawson pressure criterion achieved by nuclear reactors grew from
$3 \cdot 10^{-6}\ bar \cdot s$ in 1960 to
$3 \cdot 10^{-4}\ bar \cdot s$ in 1970 to
$2 \cdot 10^{-2}\ bar \cdot s$ in 1980 to
$4 bar \cdot s$ in 1997 \cite[p. 106]{Fus1}.
Then development of new larger reactors ran up against the lack of funding.  As of 2021, records obtained in the 1990s stand.

According to a 1997 source, ITER is supposed to be built and running no later than 2010 \cite[p.15]{BeegITER}.
A 2011 source puts ITER starting date at 2026 \cite[p.109]{Fus1}.
A 2018 source puts that date at 2035 \cite[p.65]{Fus3}.
Given that most previous predictions have turned out to be too optimistic, the starting date of ITER is unknown.
Under the best-case scenario, the first deuterium -- tritium thermonuclear power plants will appear in the 2050s \cite[p.65]{Fus3}.    In his previous work, the author has forecast that photovoltaic solar power would grow to become the World's leading energy source \cite[p.36755]{ShubovPM}.  This may also slow down development of fusion power.

Nevertheless, deuterium -- tritium fusion power is likely to play a significant role in the second half of this century.  We expect many technological improvements, such as less expensive and more durable high temperature superconductor tape, to make fusion energy cost-effective by that time.

Introduction of deuterium -- $^3$He fusion power will take place only after deuterium -- tritium fusion technology will reach maturity.  As we have mentioned in Section {\rr 2.5}, deuterium -- $^3$He reactors must have reactor criterion $\mathcal{R}_{_C}$ at least 424 times higher than similar deuterium -- tritium reactors.  It is likely that proliferation of deuterium -- $^3$He reactors would take place only in the next century.

Another major obstacle for deuterium -- $^3$He is low availability of $^3$He.
Resources of $^3$He are almost non-existent on Earth.  USA has a stockpile of about 30 $kg$ $^3$He -- a fraction of the amount needed to run a fusion reactor \cite{EarthHe3}.

Moderate deposits of $^3$He are available on the Moon.  Overall reserves of $^3$He in Lunar Regolith are estimated at 2.5 million tons \cite{MoonHe3}.  Energy content of $^3$He is about $2 \cdot 10^7$ times higher than that of coal \cite[p. 33]{MoonHe3_03}.  Hence, lunar $^3$He has energy equivalent of 50 trillion tons of coal.

Obtaining $^3$He would be difficult due to its low concentration in lunar regolith.  Helium concentration in the lunar rock is 30-40 parts per million (ppm) in terms of mass.  About 0.03\% of that helium is $^3$He \cite[p. 29]{MoonHe3_02}.  The highest weight concentration of $^3$He anywhere in lunar soil is 44 parts per billion (ppb) in terms of mass \cite{MoonHe3}.  Hence, the amount of lunar soil needed to be mined for a given amount of energy is at least the same as the amount of coal needed to be mined on Earth for the same energy.  A lunar plant would heat lunar regolith to 700 $^o$C, sort the volatiles, and export them to Earth \cite[p. 22]{MoonHe3_02}.  This would be possible only during advanced stages of Solar System colonization.

The main source of $^3$He in the Solar System is the outer planets -- Jupiter, Saturn, Uranus, and Neptune.  Atmospheres of these planets have $^3$He concentration of 3 ppm to 20 ppm.  Proposals for atmospheric mining in the outer solar system (AMOSS) are already being presented \cite{JupiterMine}.  This resource should become available during advanced stages of Solar System colonization.

\subsection{Spheromak2100}

Spheromak2100 is our concept of a deuterium -- $^3$He Tokamak to be used in Solar System exploration and colonization.  As we discuss in Subsection {\rr 6.1}, deuterium -- $^3$He3 Tokamaks are unlikely to appear within this century, hence the name reflects the probable year an early deuterium -- $^3$He3 Tokamak can be built.  Even though it is unlikely that engineers of that time will find any use for an archaic design, this design is likely to represent a minimum of performance.

Spheromak2100 represents a slight modification of GA Project 4437 developed by another team in 1996 \cite[p. 45]{Spheromak04}.  The reactor is shaped like a giant egg.  It has a height of 35.8 $m$ and diameter of 32.4 $m$.  A column 5.4 $m$ in diameter is running from top to bottom of the "egg" \cite[p. 45]{Spheromak04}.  The outer 1 $m$ of the "egg" would consist of a shell.  The shell and inner column would contain superconducting wires carrying current to provide the magnetic field inside the "egg".  The total current flowing through the central column is 194 $MA$.  The shell and inner column would also contain a neutron shield and heat rejection system composed of multiple tubes carrying cooling fluid.

Plasma would form a torus within the hollow space of Spheromak2100.  Similar to GA Project 4437 Tokamak, it has major radius $R=9.45\ m$, minor radius $a=6.75\ m$, aspect ratio $A=1.4$, and elongation $\kappa=2.5$.

GA Project 4437 Tokamak has toroidal magnetic field $B_{_T}=2.7\ Tesla$, $\beta=.61$, and operating temperature of 100 $keV$.  Its fusion power is 11 $GW$ \cite[p.45]{Spheromak04}.  So far, $\beta=0.4$ is the record demonstrated for any Tokamak \cite[p.29]{Spheromak02}.  This record has been set in 1998, and in 2021 there has not been any indication of this record being surpassed.  Thus, $\beta=0.62$, is unreasonably optimistic.    Spheromak2100 has $\beta=0.4$, toroidal magnetic field $B_{_T}=4.1\ Tesla$, and operating temperature of 70 $keV$.

At this point, we can calculate the fusion power of Spheromak2100.  In (\ref{1.04.45}), we demonstrate that the power of a Tokamak is proportional to $\beta^2\ B_{_T}^4$.  Tokamak power is also proportional to $\sigma_{_{P}}(T)$ tabulated in Table \ref{1.0T02}.  Hence, the fusion power of Spheromak2100 is
    \be
    \label{1.06.01}
    P_{_{\text{Spheromak2100}}} =
    P_{_{\text{GA Project 4437 Tokamak}}}\
    \left[
    \frac{\sigma_{_{P}}(T)\ \beta^2\ B_{_T}^4\Big|_{\text{Spheromak2100}}}
    {\sigma_{_{P}}(T)\ \beta^2\ B_{_T}^4\Big|_{\text{GA Project 4437 Tokamak}}}
    \right]=31\ GW.
    \ee
An increase in Tokamak power corresponds to an increase in heat flux at the wall.
GA Project 4437 Tokamak has been designed with maximum heat flux at the wall of 10 $MW/m^2$.  We make an optimistic assumption, that in 2100, a wall loading of 30 $MW/m^2$ would be sustainable.
The wall reflectivity of Spheromak2100 fusion chamber is $w_{_r}=0.8$.

Below, we calculate $f_{_{\text{Synchrotron}}}$ for Spheromak2100.  The values of $f_{_{\text{Synchrotron}}}$ for Spheromak2100 are calculated by (\ref{1.03.15}) and tabulated in Table \ref{1.0T13} below.  The last row is $f_{_{\gamma}}$ -- fraction of fusion power radiated away via synchrotron and Bremsstrahlung radiation.

\begin{center}
  \begin{tabular}{|l|l|l|l|l|l|l|l|l|l|l|l|}
     \hline
     Quantity & Unit & & & & & & & & \\
     \hline
     T &$keV$
     & 30 & 40   &  50   & 60   & 70   &  80  & 90  & 100  \\
     \hline
     $\sigma_{_{P}}(T)$ &
     $10^{-2}\ bar^{-1}  s^{-1}$ &
     2.54 & 3.41 & 3.80  & 3.86 &  3.73 &  3.49 &  3.21 &  2.94 \\
     \hline
     $x_{_r}\ f_{_B}$ & &
     1.32 & 0.64 & 0.41  & 0.31 &  0.25 &  0.22 &  0.2  & 0.19  \\
     \hline
     $x_{_r}\ f_{_{\text{Synchrotron}}}$ & &
     %0.14 & 0.16 & 0.21  & 0.29 &  0.38 &  0.51 &  0.68 &  0.89  \\
     0.08 & 0.09 &  0.12 & 0.16 & 0.21  & 0.28  & 0.37 &  0.49 \\
     \hline
     $x_{_r}\ f_{_{\gamma}}$ & &
     %1.46 & 0.80 & 0.63  & 0.60 &  0.63 &  0.73 &  0.88 &  1.00  \\
     1.40  & 0.73 & 0.53 & 0.47 & 0.46 & 0.50 & 0.57   & 0.68 \\
     \hline
   \end{tabular}
   \captionof{table}{Radiation losses for Spheromak2100} \label{1.0T13}
\end{center}
As we see from Table \ref{1.0T13} above, the optimal operating temperature of deuterium -- $^3$He reactor is between 60 $keV$ and 70 $keV$.

Now we calculate the value of Greenwald ratio for Spheromak2100.   From Table \ref{1.0T11}, we see that GA Project 4437 Tokamak has $R_{_G}=0.48$.  These two reactors have the same major radius $R$ and safety factor $q$.
From  (\ref{1.05.07}), it follows that $R_{_G}$ of the two reactors would be proportional to $\beta\ B_{_T}\ T^{-1}$.  Hence, Spheromak2100 has $R_{_G}=0.65$.

Now we calculate $f_{_{\text{heat}}}$ for Spheromak2100.  The fraction of fusion power which is used to heat the plasma exclusive of the power lost to Bremsstrahlung and synchrotron radiation is given in (\ref{1.04.04}).  For deuterium -- $^3$He fusion, neutronicity is 5\% \cite[p.24]{Tokamak}.  As we see from Table \ref{1.0T13}, for deuterium -- $^3$He fusion $f_{_{\gamma}}=0.46\ x_{_r}^{-1}$.  Based on the values of neutronicity and synchrotron and Bremsstrahlung power fractions presented above, we conclude that
    \be
    \label{1.06.02}
    f_{_{\text{heat}}}=0.95- 0.46\ x_{_r}^{-1}.
    \ee
for deuterium -- $^3$He plasma.
Based on Eq. (\ref{1.04.08})  and data from Table 3 and Eq. (\ref{1.06.02}), we find that for Spheromak2100, Lawson pressure criterion is
    \be
    \label{1.06.03}
    C_{_{LPI}}=\frac{6}{f_{_{\text{heat}}}\  \sigma_{_{P}}(T)}
    =\frac{6}{0.041\ bar^{-1}\ s^{-1} \big(0.95- 0.46\ x_{_r}^{-1} \big)}
    =\frac{299\ bar \cdot s}{1.94-0.94\ x_{_r}^{-1}}.
    \ee
Another source gives a value of $C_{_{LPI}} = 430\ bar \cdot s$ for deuterium -- $^3$He Lawson pressure ignition criterion \cite[p.81]{AF01}.  As we have shown in Section 1.5, $f_{_{\text{heat}}}$ for a deuterium -- $^3$He strongly depends on the reactor.  Thus, a more optimistic value given by the previous source is not a contradiction.

In Table \ref{1.0T14} below, we summarize properties of Spheromak2100.  The values of Bremsstrahlung and synchrotron radiation power are based on data in Table \ref{1.0T13}.  Normalized beta is calculated by (\ref{1.04.32}).
\begin{center}
  \begin{tabular}{|l||l|l|l|l|l|l|l|l|l|l|l|}
     \hline
     \textbf{Reactor dimensions}  & \textbf{Reactor performance} \\
     Shape: egg-like              & Fusion power: 31 $GW$        \\
     Height: 35.8 $m$             & Bremsstrahlung power: $\big(7.8\ x_{_r}^{-1}\big)\ GW$  \\
     diameter: 32.4 $m$           & Synchrotron power:    $\big(6.5\ x_{_r}^{-1}\big)\ GW$  \\
     Shell thickness: 1 $m$       & Synchrotron wall reflectivity: $w_{_r}=0.8$   \\
     Inner column radius: 2.7 $m$ & Greenwald ratio: $R_{_G}=0.65$                \\
     Plasma volume: 15,600 $m^2$  & Toroidal current: $I=202\ MA$                 \\
     Inner wall area: 3,200 $m^2$ & Beta: $\beta=0.4$                             \\
     \cline{1-1}
     \textbf{Plasma torus properties}   & Normalized beta: $\beta_{_N}=5.6$       \\
     Major radius: $R=9.45\ m$    &           \\
     Aspect ratio: $A=1.4$        &           \\
     Plasma elongation: $\kappa=2.5$ &        \\
     Plasma triangularity: $\delta=0.8$ &     \\
     Shape factor: $S_{_F}=60$    &           \\
     Toroidal field: $B_{_T}=4.1\ Tesla$ &    \\
     Plasma temperature $T=70\ keV$ &         \\
     \hline
   \end{tabular}
   \captionof{table}{Properties of Spheromak2100} \label{1.0T14}
\end{center}

\section{Conclusion}
\subsection{Summary of energy balance in fusion reactors}
In this subsection, we describe the energy balance of two typical Tokamaks -- a deuterium -- tritium reactor similar to Big ITER and a deuterium -- $^3$He Tokamak similar to Spheromak2100.  Energy balance in these reactors is representative of energy balance in all possible Tokamaks and Stellarators.

Deuterium -- tritium reactor would be running at a temperature of 20 $keV$.  From Table \ref{1.0T09} we find that operating the reactor at 15 $keV$ minimizes radiation loss.  Nevertheless, operating at higher temperature increases plasma conductivity and reduces constraints imposed by Greenwald density limit.  Deuterium -- $^3$He reactor would operate at 70 $keV$.

In Table \ref{1.0T14} below, we present the energy balance of the two aforementioned reactors.  The values of $x_{_r}\ f_{_{\gamma}}$ on Row 5 are obtained from Table \ref{1.0T09} and Table \ref{1.0T13}.  In Row 6, $P_{_{\text{Conductive}}}$ is the conductive heat loss.  The magnitude of $P_{_{\text{Conductive}}}$ is weakly dependent or independent on reactor power.
\begin{center}
  \begin{tabular}{|l|l|l|l|l|l|l|l|l|l|l|l|}
     \hline
                               & Big ITER & Spheromak2100 \\
     \hline
     Reactor type               & D -- T   & D -- $^3$He   \\
     \hline
     Operating temperature      & 20 $keV$ & 70 $keV$      \\
     \hline
     Neutronicity $f_{_n}$      & 0.8      & 0.05          \\
     \hline
     $x_{_r}\ f_{_{\gamma}}$    & 0.036    & 0.46          \\
     \hline
     $P_{_{\text{Conductive}}}$ & 20 $MW$ to 300 $MW$ & 400 $MW$ to 6 $GW$ \\
     \hline
   \end{tabular}
   \captionof{table}{Energy balance for two Tokamaks} \label{1.0T15}
\end{center}

Recall the expressions for $x_{_r}$:
    \be
    \label{1.07.01}
    x_{_r}
    \approx \left\{
    \begin{split}
    &\left[\frac{2-2\ x_{_b}}{2-x_{_b}}\right]^2
    \left[1-\frac{4 x_{_i}}{2-x_{_b}} \right]
    \qquad \text{for D -- T  fusion}
    \\
    &\big(1-x_{_b}\big)^2\ \big(1-2x_{_i}\big)
    \hskip1.98cm \text{for D -- $^3$He fusion}.
    \end{split}
    \right.
    \ee
In Eq. (\ref{1.07.01}) above, $x_{_b}$ is the proportion of the fuel burned and $x_{_i}$ is the proportion of impurities within plasma.  The fraction of fusion power which is used to heat the plasma exclusive of the power lost to Bremsstrahlung and synchrotron radiation is plotted below:
\begin{center}
\includegraphics[width=16cm,height=12cm]{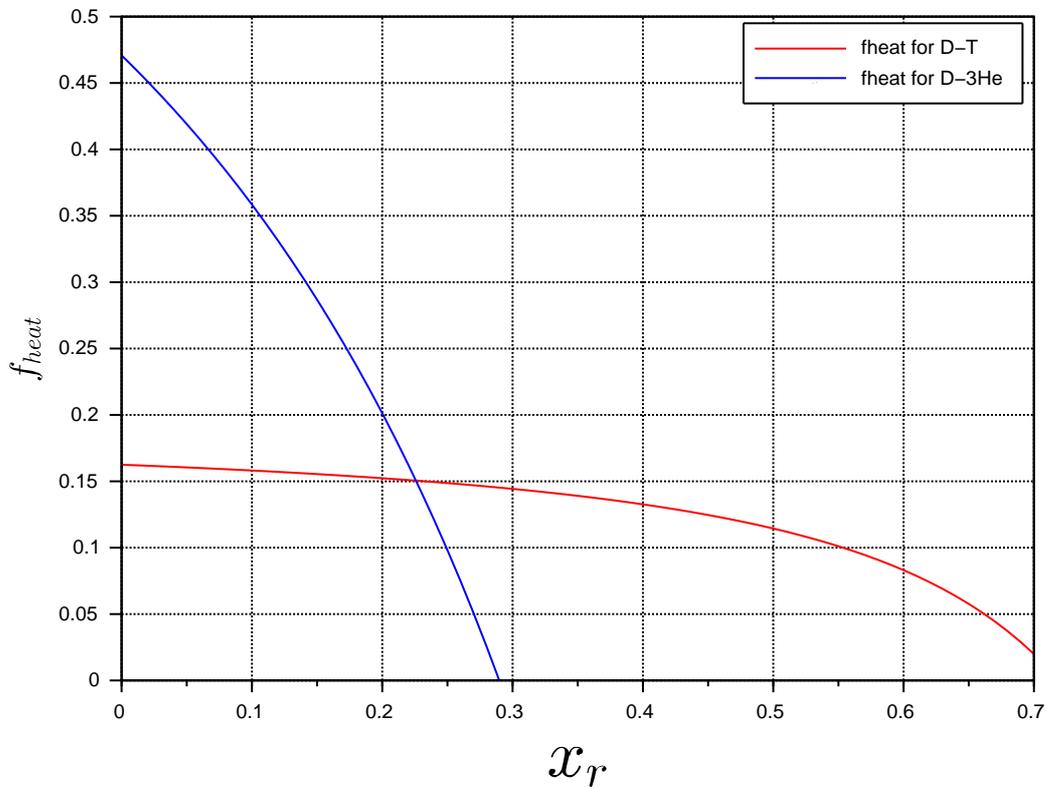}
\captionof{figure}{Tokamak cross-section \label{1.0F04}}
\end{center}
As we see from Figure \ref{1.0F04} above, the proportion of the fuel burned should not exceed 20\% for deuterium -- $^3$He fusion and 50\% for deuterium -- tritium fusion.

\subsection{Remaining problems}

The first open problem is building even one Tokamak or Stellarator with fusion power of several hundred MW to several GW.  ITER should be the first such reactor -- it would have fusion power of 500 $MW$.  Its construction cost is \$25 Billion \cite[p. 1]{LCostF1}.  We do not know when ITER will be running.  A 1997 source puts ITER starting date at 2010 \cite[p.15]{BeegITER}.  A 2018 source puts that date at 2035 \cite[p.65]{Fus3}.

The second open problem is calculating conductive power loss within fusion reactors.  In Subsection 4.2, we have mentioned several theoretical and experimental models for energy confinement time.  Each of these models has a corresponding model for conductive power loss.
Combining (\ref{1.04.37}), (\ref{1.04.39}), and (\ref{1.04.41}) we obtain the following set of models for conductive power loss:
    \be
    \label{1.07.02}
    P_{_{\text{Conductive}}} \propto
    \left\{
    \begin{split}
    &B_{_T}^{0.32} R^{0.32} A^{1.42} \beta^{1.9} q^{3} \kappa^{4.29} T^{1.32} \overline{M}^{-0.61} (\overline{Z}+1)^{1.32}
    \ \ \ \ \text{for ``IPB98(y,2)" model}
    \\
    &B_{_T}^{0.43} R^{0.43} A^{1.14} \beta^{1.74} q^{2.46} \kappa^4 T^{1.11} \overline{M}^{-0.23} (\overline{Z}+1)^{1.11}
    \ \ \text{for model from \cite{Tokamak}}
    \\
    &A\ \beta\ q^2\ \kappa^{3.5}\ T\ (\overline{Z}+1)
    \hskip5.1cm
    \text{for model from \cite{ST13}}
    \end{split}
    \right.
    \ee
More accurate power coefficients will be obtained only when Tokamaks with fusion power in hundreds $MW$ to several $GW$ are built.

The third open problem is developing a detailed understanding and tools for valid prediction of synchrotron radiation power loss.  As we have mentioned in Subsection {\rr 3.1}, different theories predict different rates of plasma energy loss by synchrotron radiation.
These results vary by up to a factor of 2 \cite[p.70]{BeegITER}.

The fourth open problem is the study of behavior of spherical Tokamaks.  What is the highest value of $\beta$ under which a spherical Tokamak can operate?  What is the most accurate model for conductive power loss in spherical Tokamaks?  In order to answer these questions, large spherical Tokamaks must be built.

\end{document}